\documentclass[preprint,12pt]{elsarticle}

\usepackage{graphicx}
\usepackage{float}
\usepackage{subfig}							% Required
\usepackage[version=3]{mhchem} 	% Required
\usepackage{epstopdf}						% Required
\usepackage{array}							% Required
\usepackage[displaymath, mathlines]{lineno}

\graphicspath{{figures/}}

\journal{Acta Materialia}

\begin{document}

\begin{frontmatter} 

\title{Isothermal Annealing of Shocked Zirconium: Stability of the Two-phase $\alpha/\omega$ Microstructure}

\author[l1]{T.S.E. Low}
\author[l3]{D.W. Brown}
\author[l4]{B.A. Welk}
\author[l3]{E.K. Cerreta}
\author[l5]{J.S. Okasinski}
\author[l1,l6]{S.R. Niezgoda\corref{cor1}} 
\cortext[cor1]{Corresponding Author}
\ead{niezgoda.6@osu.edu}

\address[l1]{Department of Materials Science and Engineering, The Ohio State University, Columbus, OH 43210, USA}
\address[l3]{Materials Science and Technology Division, Los Alamos National Laboratory, Los Alamos, NM 87545, USA}
\address[l4]{Center for the Accelerated Maturation of Materials, The Ohio State University, Columbus, OH 43210, USA}
\address[l5] {Advanced Photon Source, Argonne National Laboratory, Argonne, IL 60439, USA}
\address[l6]{Department of Mechanical and Aerospace Engineering, The Ohio State University, Columbus, OH 43210, USA}
\begin{abstract}
Under high pressure conditions, Zr undergoes a phase transformation from its ambient equilibrium hexagonal close packed $\alpha$ phase to hexagonal $\omega$ phase. Upon returning to ambient conditions, the material displays hysteretic behavior, retaining a significant amount of metastable $\omega$ phase. This study presents an in-situ synchrotron X-ray diffraction analysis of Zr samples shock-loaded to compressive peak stresses of 8 and 10.5 GPa and then annealed at temperatures of 443, 463, 483, and 503K. The evolution of the $\alpha$ phase volume fraction was tracked quantitatively, and the dislocation densities in both phases were tracked qualitatively during annealing. Upon heating, the reverse transformation of $\omega\to\alpha$ does not go to completion, but instead reaches a new metastable state. The initial rate of transformation is faster at higher temperatures. Samples shock-loaded to higher peak pressures experienced higher initial transformation rates and more extensive transformation. Dislocation content in both phases was observed to be high in the as-shocked samples. Annealing the samples reduces the dislocation content in both phases, with the reduction being lesser in the $\omega$ phase, leading to the postulation that transformation from $\omega\to\alpha$ is restricted by the pinning effect of dislocation structures within the $\omega$ phase. Electron backscatter diffraction analysis affirmed that the expected $(0\;0\;0\;1)_\alpha\parallel(1\;0\;\overline{1}\;1)_\omega$ and $[1\;0\;\overline{1}\;0]_\alpha\parallel[1\;1\;\overline{2}\;\overline{3}]_\omega$ orientation relationship is maintained during nucleation and growth of the $\alpha$ phase during the annealing. 
\end{abstract}

\begin{keyword}
Zirconium
\sep Synchrotron Diffraction
\sep High Pressure
\sep Phase Transformation
\sep Annealing
\end{keyword}

\end{frontmatter}
%\linenumbers
\section{Introduction}
\label{Section_Intro}

At ambient temperatures and pressures, zirconium is thermodynamically stable as a single phase, $\alpha$, hexagonal close packed (HCP) crystal ($P6_3/mmc$, $c/a=1.593$). Under high pressures, Zr and other transition HCP metals such as Ti and Hf  undergo a phase transformation from the $\alpha$ phase to the simple hexagonal $\omega$ phase ($P6/mmm$, $c/a=0.623$). The $\omega$ phase has been shown to display strong hysteretic behavior under both static and shock loading, with the high-pressure $\omega$ phase being retained after pressure is released \cite{jamieson, Vohra, Rabinkin, kutsar, greeff, jyoti2008}. By shock-loading or deforming Zr to peak pressures above 7 GPa, as much as 80\% of the $\omega$ phase has been retained after subsequent unloading to ambient pressures \cite{cerreta2005influence, cerreta2012influence, Cerreta20137712}, and metastable nano-grained  $\omega$ Zr and Ti has been produced by high-pressure torsion \cite{zhilyaev2010phase,edalati2009allotropic,srinivasarao2011orientation,perez2008bulk,xia1991temperature}. This is in contrast to the high-pressure phase transformation in Fe, where the $\varepsilon$ phase is not quenchable \cite{bancroft}.

Modeling the dynamic deformation behavior of Zr and Ti necessitates a robust understanding of the coupling between the mechanisms of plastic deformation at high strain rates and the $\alpha/\omega$ phase transformation at high pressures. An important part of this understanding is the characterization of the hysteresis of the transformation and the stability of the two-phase microstructure. Cultivating this understanding is central to elucidating the kinetics of the transformation that are applicable to dynamic conditions far from thermodynamic equilibrium. The current work focuses on understanding the mechanisms for arresting the reverse transformation post-shock and the effects of temperature and peak shock pressure on the reverse transformation during subsequent annealing. 

Investigations of the $\alpha \rightarrow \omega$ transformation in Zr and Ti have largely been focused on determining the equilibrium pressure, crystallography, and resulting mechanical properties and electronic structure of the $\omega$ phase \cite{Rabinkin, greeff2004modeling, cerreta2003shock, trinkle2005systematic}. Multiple orientation relationships have been observed each with one or more proposed mechanisms or transformation pathways \cite{Rabinkin, jyoti2008, usikov1973orientation, SarathKumarMenon1982717, song1995microscopic, Gupta19851167, trinkle2003}. Investigations of the kinetics of the phase transformation in Zr have largely been limited to the forward $\alpha \rightarrow \omega$, with little found in the literature concerning the reverse transformation of $\omega\rightarrow\alpha$ or the stability of the dual phase microstructure. Recently, Brown et al. examined the stability of the $\omega$ phase in shock-loaded Zr in-situ at a constant heating rate of 3K/min from room temperature to $\sim$620K \cite{Brown2014383}. Based on the observation of large initial dislocation densities (post-shock) in both the $\alpha$ and $\omega$ phases and subsequent decrease in dislocation content in $\omega$ preceding the reverse transformation, it was speculated that the hysteresis in shocked-samples is due, at least in part, to the high concentration of defects in the $\omega$ phase retarding the transformation and preventing the system from returning to  thermodynamic equilibrium after completion of the shock \cite{Brown2014383}.  

In the current work, X-ray diffraction techniques were used to characterize the microstructural evolution of shock loaded Zr specimens during isothermal annealing at low homologous temperatures. More specifically, the measurements in this study include quantitative measurements of the growth of the $\alpha$ phase and the subsequent stability of the two phase microstructures.  Dislocation density trends in both phases were monitored through the evolution of the root mean square (RMS) strain $\varepsilon_{rms}$ and related to the phase transformation. The morphology of the microstructure was examined by selected electron backscatter diffraction (EBSD) studies on the as-shocked and partially annealed specimens. 

\section{Experimental Methods}
\label{sec:experimental_methods}
\subsection{Sample preparation}
All samples were prepared from a high-purity crystal bar Zr ($<$100 ppm impurities) which was upset forged, clock rolled, and annealed at 823K for 1 hour \cite{escobedo2012influence}, producing a plate with a homogenous and fully recrystallized microstructure, having an average grain size of 15-20 $\mu m$ \cite{escobedo2012influence}. The plate exhibits an in-plane isotropic crystallographic texture with a strong basal component ($>$8 times uniform random distribution) nearly aligned with the normal or through-thickness direction (TT) direction of the plate, with prism planes uniformly distributed about the in-plane directions of the plate. 5 mm thick, 25 mm diameter Zr disks were electrodischarge machined (EDM'ed) from the rolled plate with the symmetry axis of the disk parallel to the rolling normal (TT) direction for the shock loading. The specimens were then tightly fit into target assembles specifically designed for shock loading/unloading experiments \cite{grayiiigt1993, Koller}. The targets were impacted by 2.5 mm thick Zr flyer plates accelerated to velocities of 640 or 835 m/s, resulting in peak compressive stresses of 8 or 10.5 GPa respectively on the Zr samples \cite{Cerreta20137712}.

From the soft-recovered \cite{grayiiigt1993, ASM2000} shocked specimens, 1 mm thick $\times$ 3 mm diameter discs were EDM'ed for X-ray and microscopy examination. The disks were machined such that the disk axis was orthogonal to the shock direction and parallel to the in-plane direction of the original plate. The samples were cold mounted and prepared for microscopy by grinding with 2400 grit aluminum oxide paper and chemically polished with a solution of \ce{45H_2O : 45HNO_3 : 10HF} to remove any surface damage caused by machining. The examined surface corresponded to the cross-section of the X-ray disks and contained both the shock direction and an in-plane sample direction. The same sample preparation is applied to samples that are subjected to heating during the in-situ X-ray diffraction experiments.

\subsection{X-ray diffraction measurement} 
\label{sec:x_ray_measurement}
The in-situ heating measurements were completed on the 1ID-C beam line at the Advanced Photon Source (APS), Argonne National Laboratory \cite{Haeffner2005120}. A Cu sample fixture was constructed for the annealing experiments to ensure uniformity of temperature. The sample holder was designed to facilitate the simultaneous annealing and characterization of two Zr samples, one subjected to 8 GPa shock pressure and the other 10.5 GPa shock pressure. The simultaneous heating of the two samples a) ensures that for a given annealing temperature, each sample is exposed to nearly identical thermal environments and b) greatly reduces the experimental time required to perform the experiment as the microstructural kinetics are slow compared to the collection time of  diffraction patterns. Two X-ray through holes of diameter 2 mm were drilled in the 5 mm thick Cu plate, separated by 10 mm on-center. A 2 mm thick copper cover plate with two 3.3 mm diameter $\times$ 0.75 mm deep recesses was attached to the larger plate with spring clamps. A Zr disk sample was placed in each recess, with one recess containing a 8 GPa loaded sample, and the other containing a 10.5 GPa loaded sample. The spring clamps were tightened sufficiently to maintain contact and ensure good thermal conductivity, but not excessive so as to cause significant unknown stresses on the sample during heating. Since the Zr disks were slightly thicker than the recesses in the Cu plate, direct thermal contact with the Cu plate on both faces was ensured. The loaded sample holder was mounted on an MTS servo-hydraulic load frame in order to translate the two samples into and out of the beam in an alternating fashion. The samples and holder were heated in a focused optical furnace, and the temperature was manually controlled by monitoring reference thermocouples embedded in the thicker Cu plate. The Cu fixture provided an additional measurement of the temperature in contact with each sample. Given the well characterized coefficient of thermal expansion of Cu \cite{touloukian1975thermal}, passing the X-ray through the thin Cu backing plate allowed for accurate tracking of the temperature change in the samples from the change in Cu peak positions.

The 1ID beam line utilizes a monochromated beam from the standard APS undulator and double-crystal Laue monochromator designed specifically for high-energy X-rays  \cite{shastri2002cryogenically}. The incident beam ($E=86$ keV), was masked to a 200 $\mu$m $\times$ 200 $\mu$m cross-section before impinging on the sample/fixture parallel to the cylinder axis of the samples. The X-ray beam penetrated 1.25 mm of the Cu backing plate as well as the Zr sample and was diffracted onto a two-dimensional GE 41RT detector with 2048$\times$2048 pixels (0.200 mm $\times$ 0.200 mm pixel size) roughly centered on the straight-through beam. The sample to detector distance is approximately 1500 mm, enabling $2\theta$ coverage of roughly $\pm 7^\circ$ or an approximate d-spacing from 1.0 to 3.5\AA. The angular coverage of the detector was sufficient to collect up to 5 complete diffraction rings for each phase.

During the in-situ annealing, the temperature was ramped as fast as possible ($\sim1.5$ K/s) from room temperature to the annealing temperature while avoiding significant overshoot. Samples shocked at both 8 and 10.5 GPa were annealed at 443, 463, 483, and 503K (reference thermocouple temperature). Samples were held at temperature until visual inspection of the collected diffraction patterns showed no change over a 15 minute increment. The annealing times varied with temperature from $2 \times 10^3$ to $>2\times10^4$ seconds. The 503K sample (10.5 GPa) was then further heated at a higher temperature 773K for 30 minutes to produce a fully annealed sample with no measurable retained $\omega$ phase. During characterization of the 463K samples, there was an unplanned X-ray beam loss for approximately 25 minutes. During this time the samples temperatures were maintained and data collection resumed upon beam restoration. 

\subsection{X-ray data analysis}
Each 2-D diffraction image was binned into 24-$15^\circ$ intervals in the azimuthal angle, $\eta$ about the through beam using Fit2D \cite{hammersley1997fit2d} to make individual 1-D diffraction patterns with diffraction vectors nearly transverse to the incident beam ($\approx 2^\circ$ off) and transverse to the cylinder axis of the sample. As the samples were cut with their cylinder axis transverse to the shock direction, the 24 diffraction vectors sample orientations from the shock direction (TT) to an IP direction.  

Whole pattern Rietveld refinements of the 1-D diffraction patterns were accomplished using GSAS software developed at Los Alamos National Laboratory \cite{larson1994gsas}. The data was analyzed to determine the volume fractions, lattice parameters, peak variances of the $\alpha$ and $\omega$ phases, and the lattice parameter of the Cu backing plate in contact with the samples.  Single diffraction peaks were analyzed for peak position, integrated intensity and peak width using the Rawplot subroutine of GSAS. The peaks were fitted with a pseudo-Voigt peak profile. The automated routines APSrunrep and APSspf \cite{clausen2003smartsware} which call GSAS subroutines, were used to enable the analysis of the tens of thousands of diffraction patterns. Quoted uncertainties are based off the estimated standard deviations returned by GSAS.  Examples of integrated 1D diffraction patterns in the as-shocked and fully annealed states have been shown previously in Brown et al. \cite{Brown2014383}. Little peak overlap between the two phases is present, allowing five unique peaks from each phase to be analyzed. Both the Rietveld and single peak fits are robust and the results are consistent internally and with previous studies \cite{Brown2014383}.

The thermal strains in the Cu plate, in conjunction with the known thermal expansion of Cu, were used to more accurately determine the temperature of the sample. In general, the temperature of the Cu plate in contact with the samples (taken to be the accurate measure of sample temperature) was in good agreement (within $\pm 2$K) with the reference thermocouples. Additionally the temperature difference between the two samples in the holder was $\le 2$K.

Information concerning the defect state of the microstructure can be estimated from the diffraction peak breadth and variance, both of which are determined for the $\alpha$ and $\omega$ phases by the Rietveld refinements in GSAS. The diffraction peak variance is frequently assumed to be a sum of contributions from the RMS strain, $\varepsilon_{rms}$, which is associated with dislocations and finite crystallite domain or sub grain size (although other microstructural effects can also affect the peak breadth). 

In this work we will use $\varepsilon_{rms}$ as a qualitative measure of dislocation density in the $\alpha$ and $\omega$ phase. In the case of a dislocated crystal, $\varepsilon_{rms}$ depends on the size of the integration distance used to calculate the average \cite{aqua1966separation, wilkens1970determination, leineweber2010notes}. For the purpose of assessing trends and qualitative measurement, this dependence can be neglected and it can be assumed with reasonable accuracy that $\varepsilon_{rms}$ is proportional to the dislocation density in each phase \cite{Brown2014383}. Unfortunately, quantitative measures such as diffraction line profile analysis are not applicable due to the limited knowledge of deformation in the $\omega$ phase (knowledge of the active slip systems and Burger's vectors are lacking) \cite{Brown2014383}. In previous work, Brown et al. performed quantitative line profile analysis on the $\alpha$ phase during constant rate heating experiments and found that the $\varepsilon_{rms}$ scaled well with the computed dislocation density and that trends in defect density evolution could be easily discerned from the RMS microstrain \cite{Brown2014383}. 

Additionally the crystallographic texture of the as-shocked and fully annealed samples was measured by rotating the samples $\pm 45^\circ$ in $\Delta \chi =15^\circ$ increments in the incident X-ray beam. For each $\chi$ angle, the data was binned in into 24 azimuthal bins and the complete set of histograms were refined simultaneously with MAUD using the Rietveld method and E-WIMV representation of the orientation distribution function with 10${}^\circ$ resolution \cite{lutterotti2007rietveld}. The texture analysis, performed largely for validation against  previously reported experiments, showed no new findings \cite{Brown2014383}. The as shocked $\alpha$ and $\omega$ transformation textures were found to be nearly uniform ($<2$ time random), and the fully annealed $\alpha$ texture was also nearly random indicating no preferential variant selection during nucleation and growth. To be succinct, the reader is referred to \cite{Brown2014383} for pole figures.  

\subsection{Electron Backscatter Diffraction}
The samples were examined in an FEI XL-30 ESEM equipped with a TSL DigiView EBSD camera. EBSD was performed on both the 8 GPa and 10.5 GPa samples, however the 10.5 GPa samples produced diffraction patterns of poor quality with exceptionally noisy results.  Although initial analysis of the 10.5 GPa samples appears to be in qualitative agreement with the results reported for the 8 GPa samples, they will not be shown until higher quality results can be collected. Post collection, the 8 GPa data was moderately cleaned up using the phase correlation clean up routine in the TSL OIM Analysis software.

\begin{figure*}[htb!]
\begin{center}

\begin{tabular}{cc}
\subfloat[8 GPa]{
\includegraphics[scale=0.31,keepaspectratio=true]{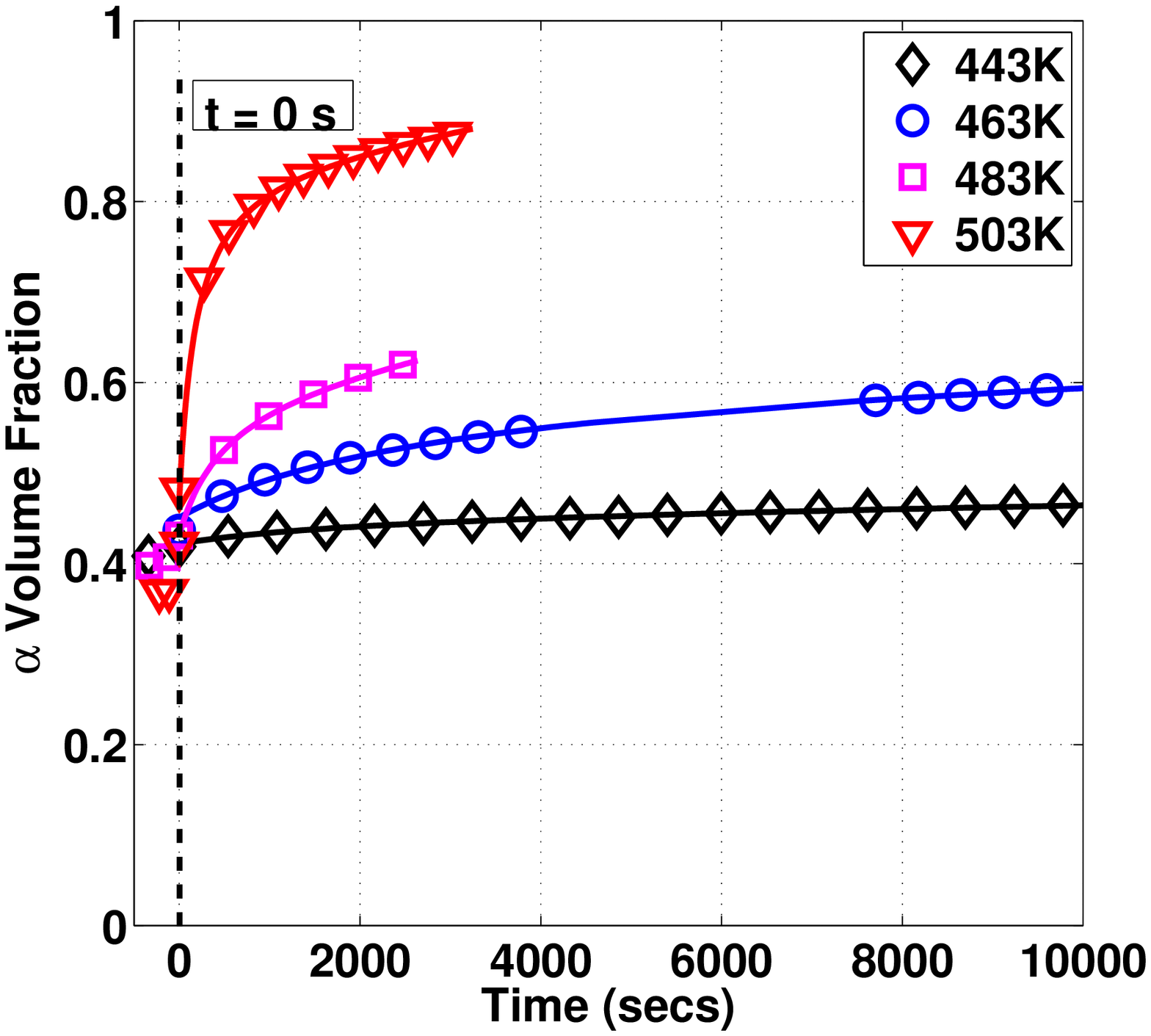}\label{fig:alpha_evolution_1_a}
} 
\subfloat[10.5 GPa]{
\includegraphics[scale=0.31,keepaspectratio=true]{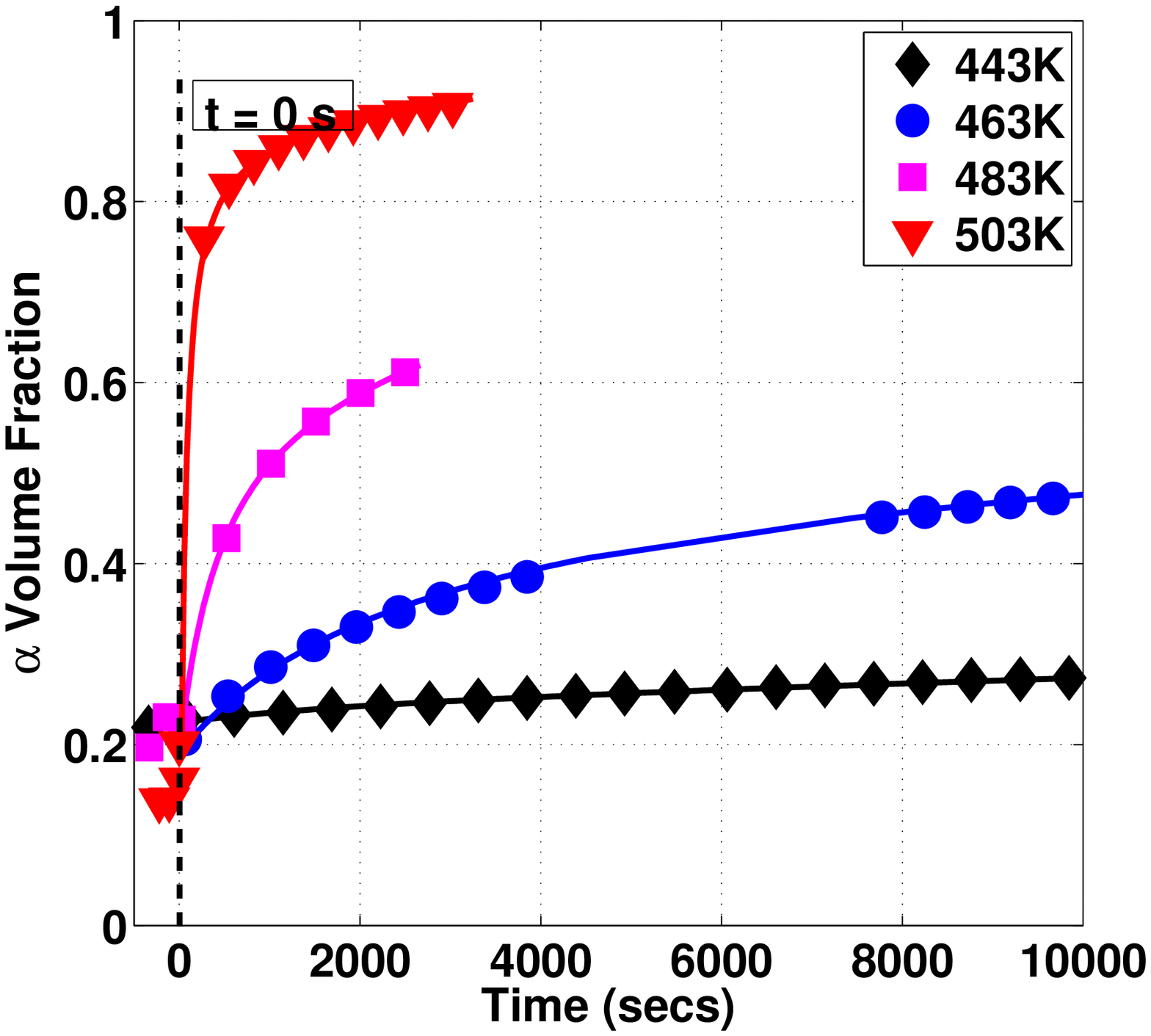}
\label{fig:alpha_evolution_1_b}
}

\end{tabular}
\end{center}
\caption{The evolution of the $\alpha$ volume fractions during isothermal heating of the material shocked to (\protect\subref*{fig:alpha_evolution_1_a}) 8 GPa and (\protect\subref*{fig:alpha_evolution_1_b}) 10.5 GPa peak pressures as a function of time at varying temperatures.  Markers indicate volume fractions from X-ray analysis, solid lines indicate 2nd order rational function fit ($(c_1 t^2+ c_2t+c_3)/(t+c_4) $) to the experimental data. For visual clarity, markers are only shown every 10 or 15 data points. The large gap between markers in the 463K experimental data is due to the X-ray beam loss described in section (\protect\ref{sec:x_ray_measurement}). The time $t=0$s corresponds to when the sample reaches 95$\%$ of the target temperature, and negative time corresponds to the heat-up phase.} 
\label{fig:alpha_evolution_0}
\end{figure*}
 
\section{Results} 

\begin{figure*}[htb!]

\begin{center}
\begin{tabular}{cc}
\subfloat[245C Spectra at t = 0 Fs]{
\includegraphics[scale=0.31,keepaspectratio=true]{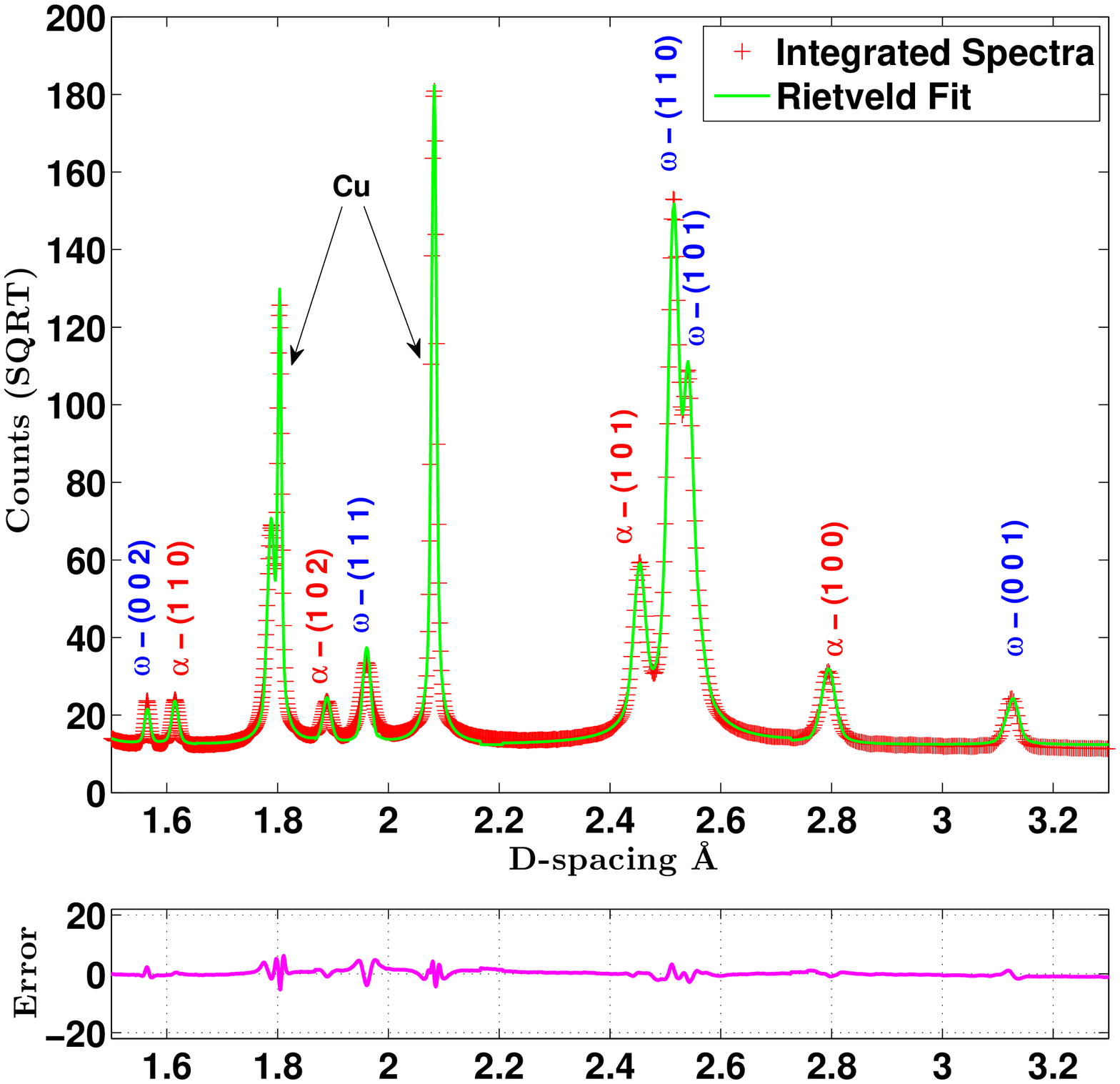}\label{fig:spectra_init}
}

\subfloat[245C Spectra at t = 3523 s]{
\includegraphics[scale=0.31,keepaspectratio=true]{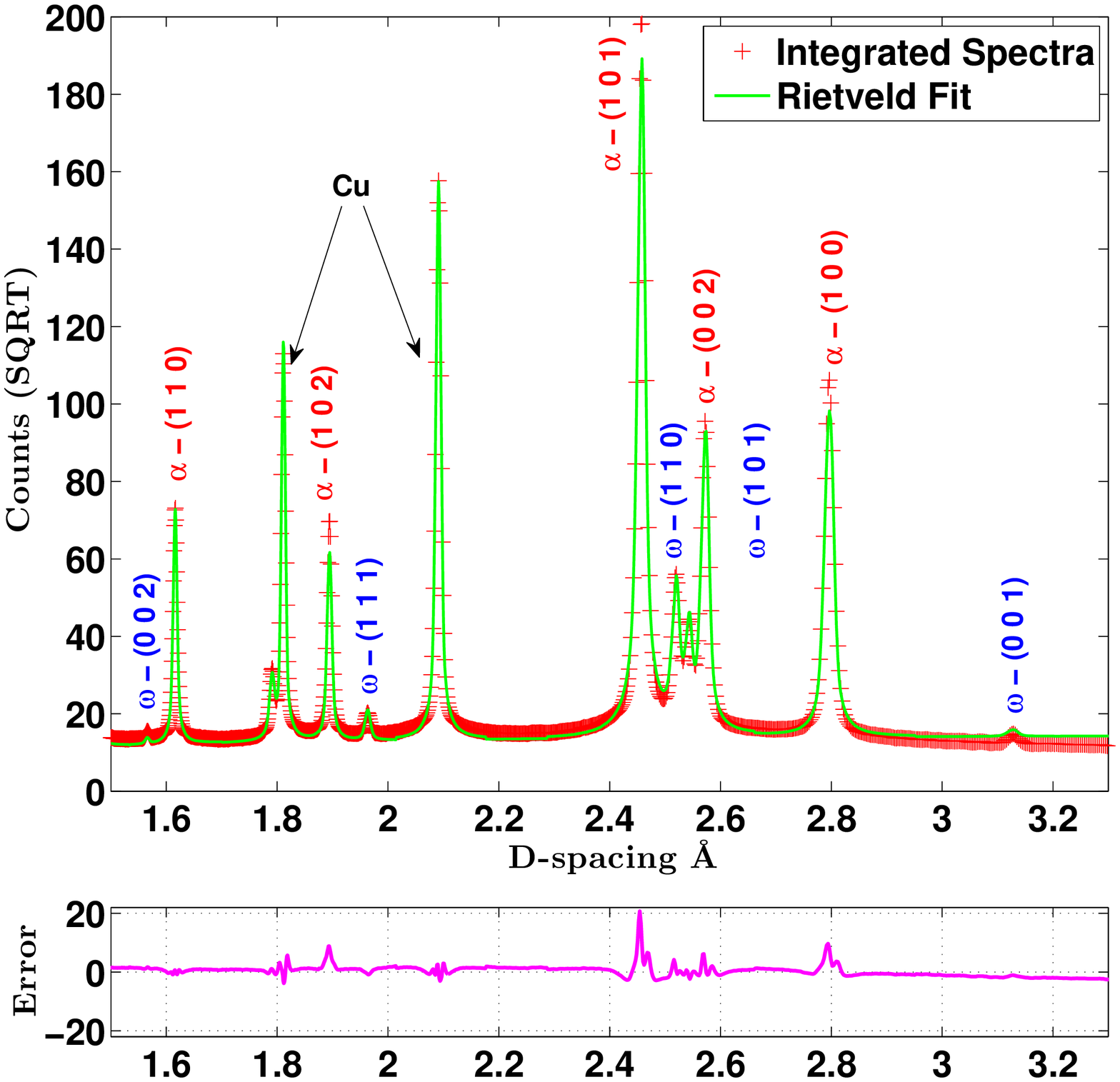}\label{fig:spectra_final}
}
\end{tabular}
\end{center}

\caption{The integrated X-ray spectra and applied Rietveld fittings for 10 GPa samples subjected to 230C heating (\protect\subref*{fig:spectra_init}) prior to heating and (\protect\subref*{fig:spectra_final}) after heating for 3523 secs. Note: The Y-axis intensity is square-root of the original intensity.}
\label{fig:spectras}
\end{figure*}

\subsection{In-situ X-ray diffraction experiments}
After shock loading, the initial $\alpha$ volume fraction for the material shocked to 8 and 10.5 GPa is $0.39\pm0.02$ and $0.18\pm0.03$ respectively. This is in good agreement with previous studies on the same material, although with slightly higher sample to sample variance \cite{Brown2014383}. $t=0$ s is taken as the time when the temperature reaches 95$\%$ of the target annealing temperature. Comparing the initial volume fractions shown in Figures \ref{fig:alpha_evolution_0}\subref*{fig:alpha_evolution_1_a} and \ref{fig:alpha_evolution_0}\subref*{fig:alpha_evolution_1_b} with those reported above, it is seen that some degree of phase-transformation occurred during the temperature ramp up. The evolution of the $\alpha$ phase fraction in both samples follow the same general trend during the isothermal hold. The initial rates of transformation (represented by the slopes of the curves in Figure \ref{fig:alpha_evolution_0}\subref*{fig:alpha_evolution_1_a} and \ref{fig:alpha_evolution_0}\subref*{fig:alpha_evolution_1_b}) are relatively high at sufficiently high temperatures but decrease with time. The $\alpha$-phase volume fraction grows and saturates at a steady state level that is less than one, instead reaching another metastable state containing both $\alpha$ and $\omega$ phases. The reader can further verify this in Figure \ref{fig:spectras} which displays the integrated spectra and corresponding Rietveld fittings for 10 GPa samples heated at 230C prior to and after the heating. the Both the initial transformation rate and extent of the $\omega\to\alpha$ transformation are strongly temperature dependent, with both increasing with temperature.

\begin{figure*}[htb!]
\begin{center}

\begin{tabular}{cc}
\subfloat[443K]{
\includegraphics[scale=0.31,keepaspectratio=true]{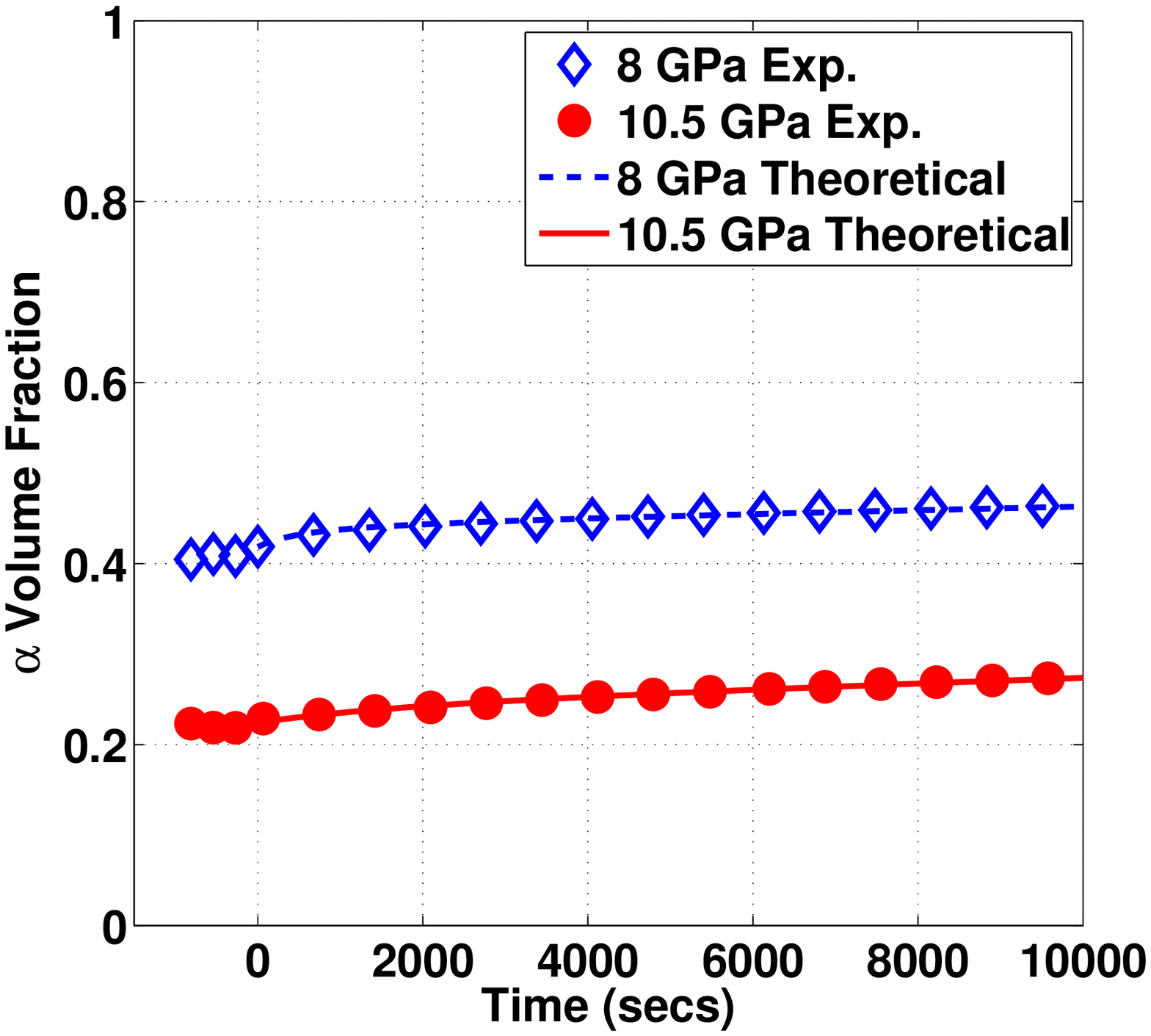}
\label{fig:alpha_evolution_2_a}
}
\subfloat[463K]{
\includegraphics[scale=0.31,keepaspectratio=true]{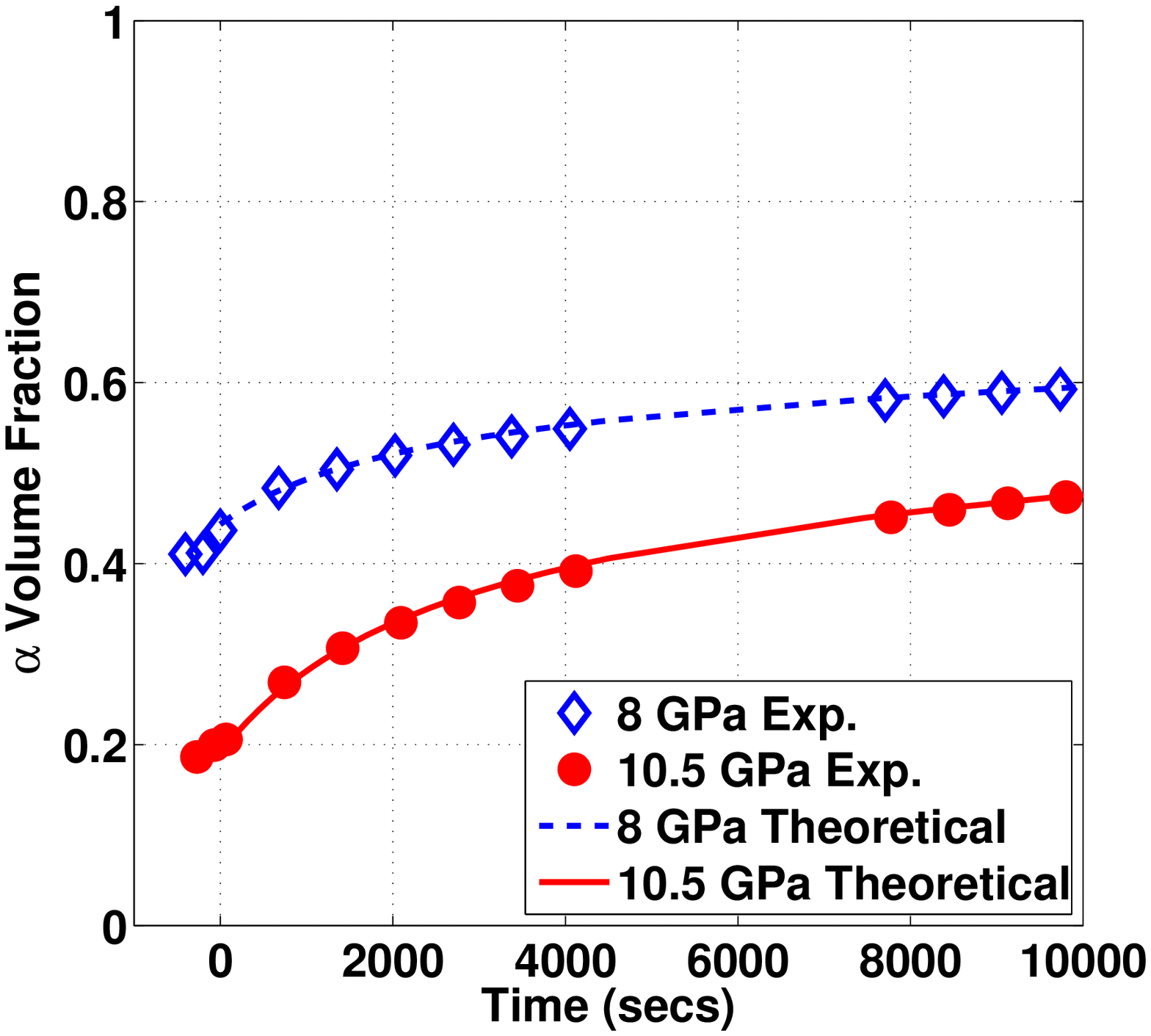}
\label{fig:alpha_evolution_2_b}
} \cr
\subfloat[483K]{
\includegraphics[scale=0.31,keepaspectratio=true]{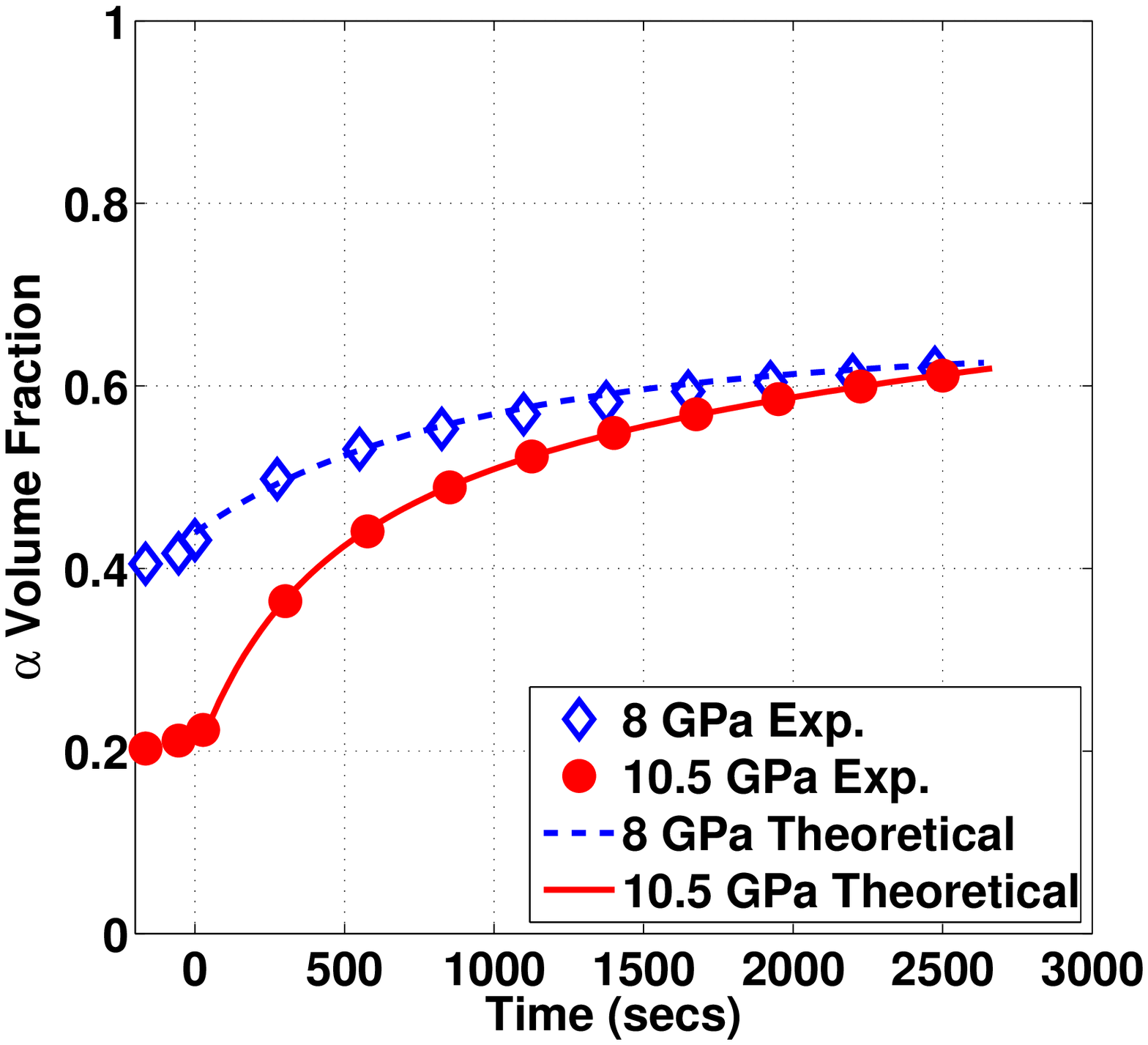}
\label{fig:alpha_evolution_2_c}
} 

\subfloat[503K]{
\includegraphics[scale=0.31,keepaspectratio=true]{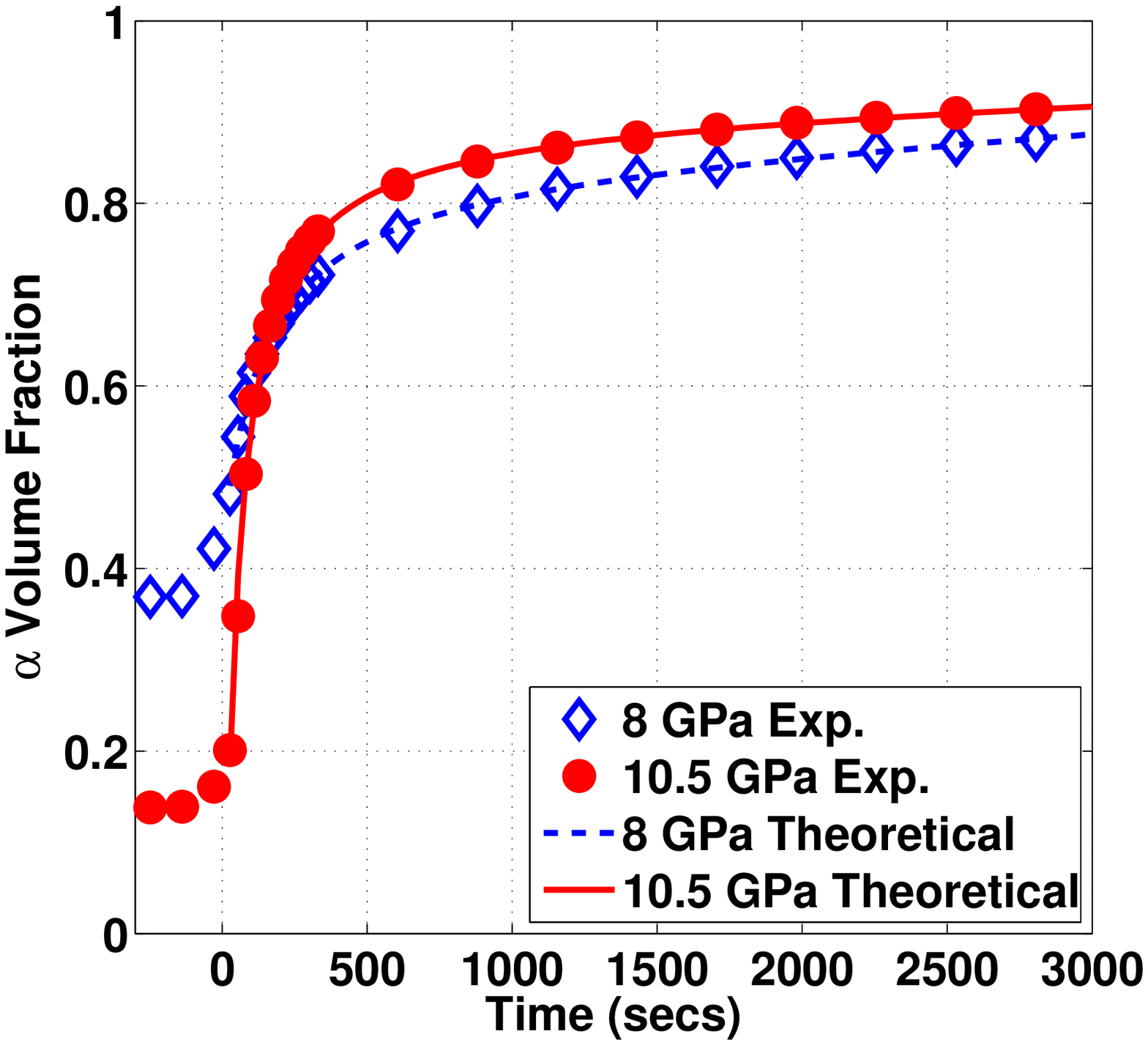}
\label{fig:alpha_evolution_2_d}
}
\end{tabular}
\end{center}
\caption{Comparing the effects of peak pressure on evolution at a fixed temperature of (\protect\subref*{fig:alpha_evolution_2_a}) 443K, (\protect\subref*{fig:alpha_evolution_2_b}) 463K,(\protect\subref*{fig:alpha_evolution_2_c}) 483K, (\protect\subref*{fig:alpha_evolution_2_d}) 503K. Different time scales are applied here.} 
\label{fig:alpha_evolution_1}
\end{figure*}

\begin{figure}[htb!]
\begin{center}
\includegraphics[scale=0.31,keepaspectratio=true]{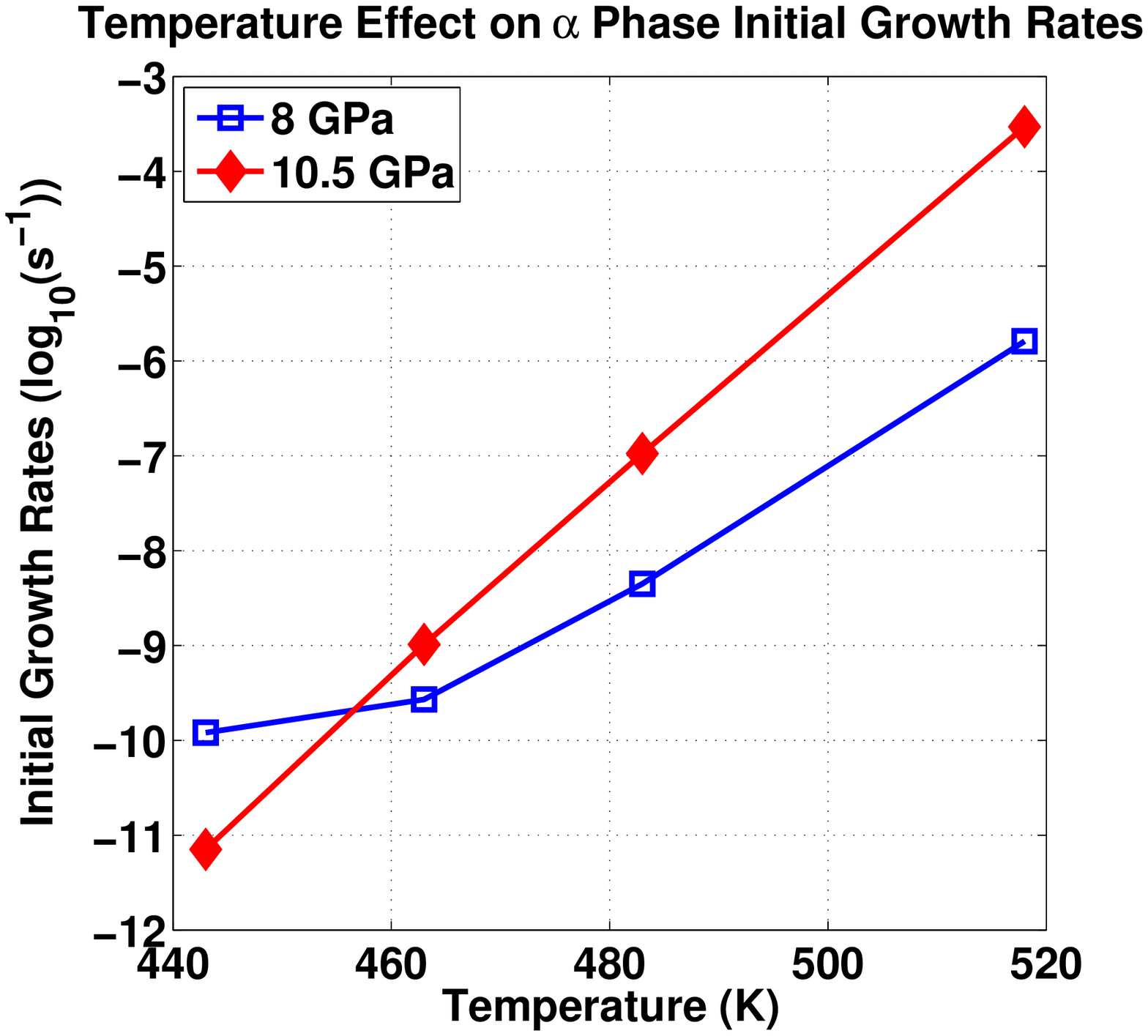}
\end{center}
\caption{Semilog plot showing initial $\alpha$ fraction growth rates with respect to temperature.}
\label{fig:init_growth}
\end{figure}

Interestingly, the transformation rate is also a function of the post-shock condition of the material. The reverse transformation from $\omega$ to $\alpha$ occurs more rapidly in the 10.5 GPa shocked sample in contrast with the 8 GPa shocked sample. Figures \ref{fig:alpha_evolution_1}\subref*{fig:alpha_evolution_2_c} and \ref{fig:alpha_evolution_1}\subref*{fig:alpha_evolution_2_d} illustrate the described effect of the pre-shocked peak pressures on the $\omega\to\alpha$ transformation. At higher temperatures, the $\alpha$ concentration in the two samples crossover, and the final steady-state $\alpha$ concentration is higher in the 10.5 GPa shocked material, despite initially having a lower $\alpha$ volume fraction. Given sufficient time, the transformation stabilizes or saturates at a higher $\alpha$ fraction than the 8 GPa sample. Figure \ref{fig:alpha_evolution_1}\subref*{fig:alpha_evolution_2_d} also highlights the degree of transformation which occurred during the temperature ramp to the final annealing temperature. Despite ramping as quickly as the experimental setup allowed, changes in volume fraction of up to 0.05 (for the 503K samples) were observed during the temperature ramp. 

\begin{figure*}[htb!]
\begin{center}

\begin{tabular}{cc}

\subfloat[$\varepsilon_{rms}$ in $\alpha$ phase (8GPa)]{
\includegraphics[scale=0.31,keepaspectratio=true]{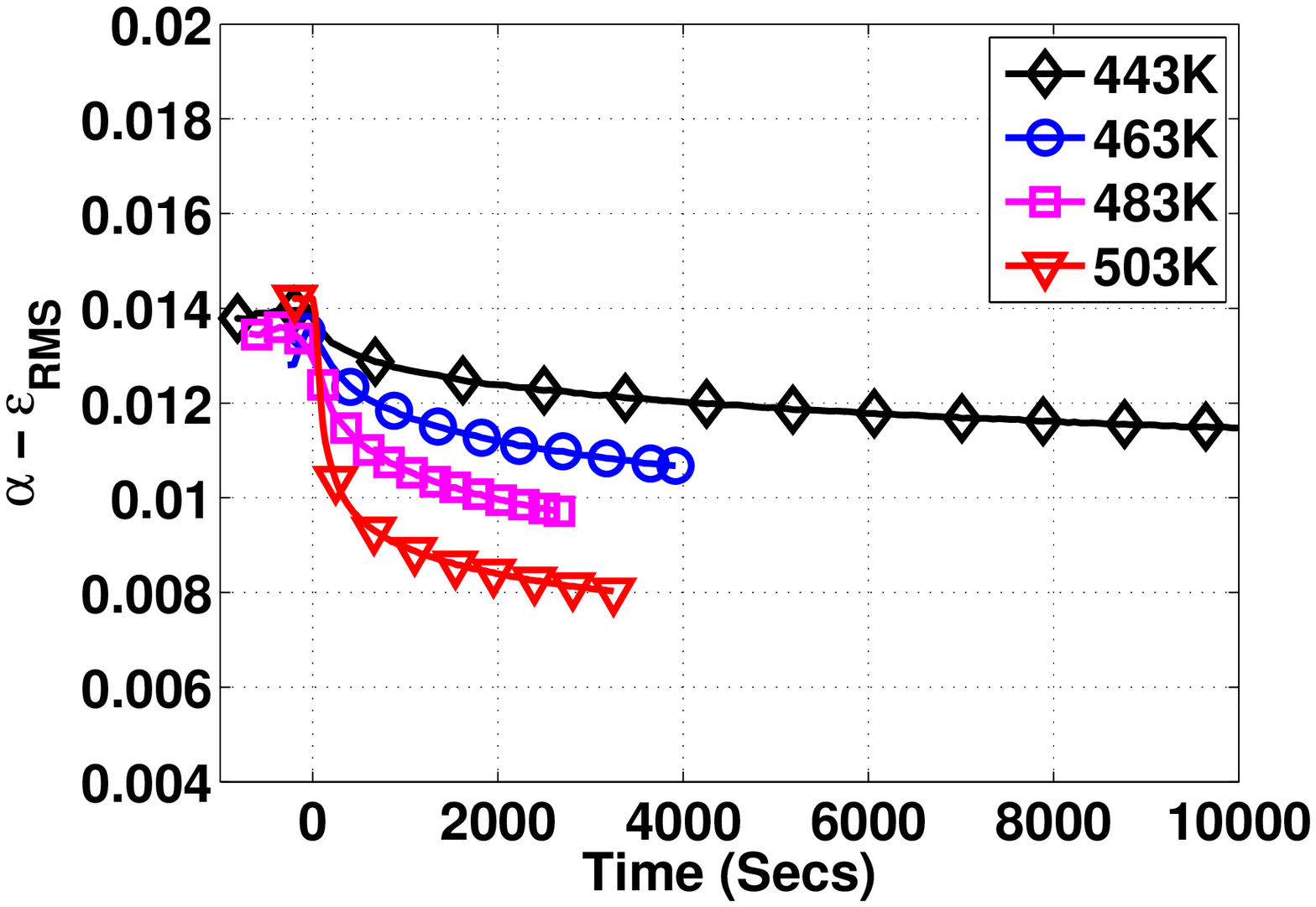}
\label{fig:rms1_a}
} 

\subfloat[$\varepsilon_{rms}$ in $\omega$ phase (8GPa)]{
\includegraphics[scale=0.31,keepaspectratio=true]{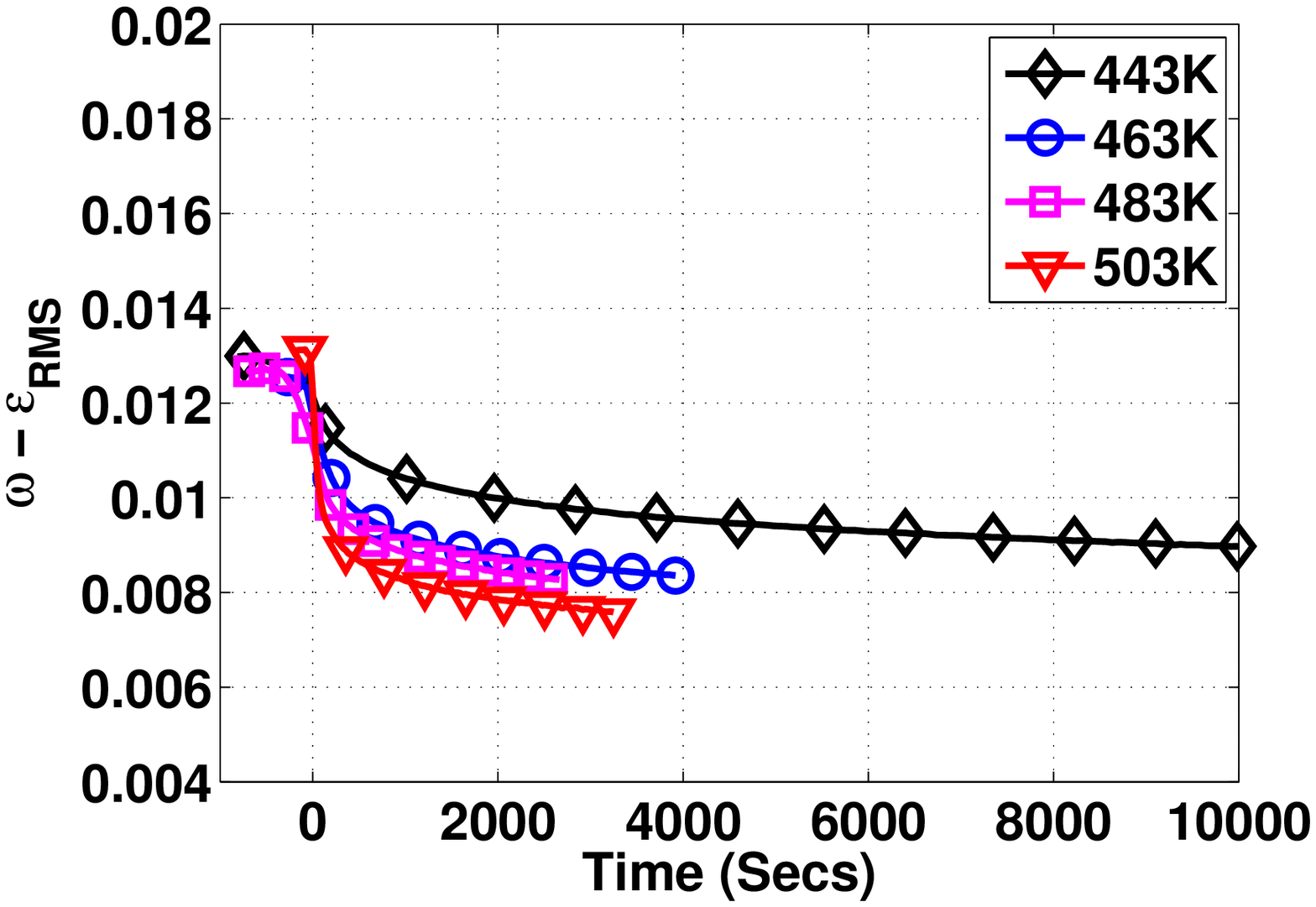}
\label{fig:rms1_b}
} 
\cr
\subfloat[$\varepsilon_{rms}$ in $\alpha$ phase (10.5GPa)]{
\includegraphics[scale=0.31,keepaspectratio=true]{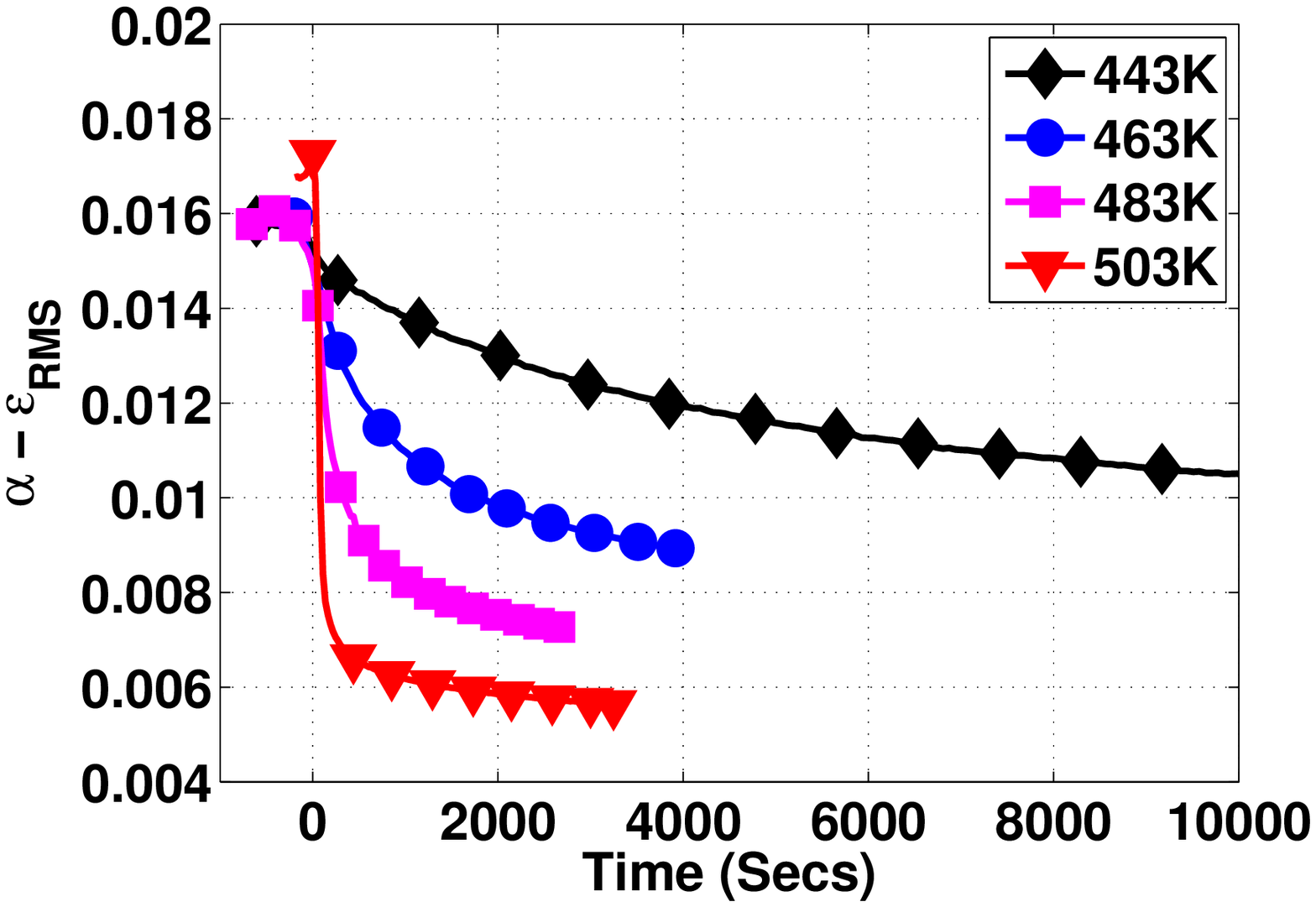}
\label{fig:rms1_c}
} 
\subfloat[$\varepsilon_{rms}$ in $\omega$ phase (10.5GPa)]{
\includegraphics[scale=0.31,keepaspectratio=true]{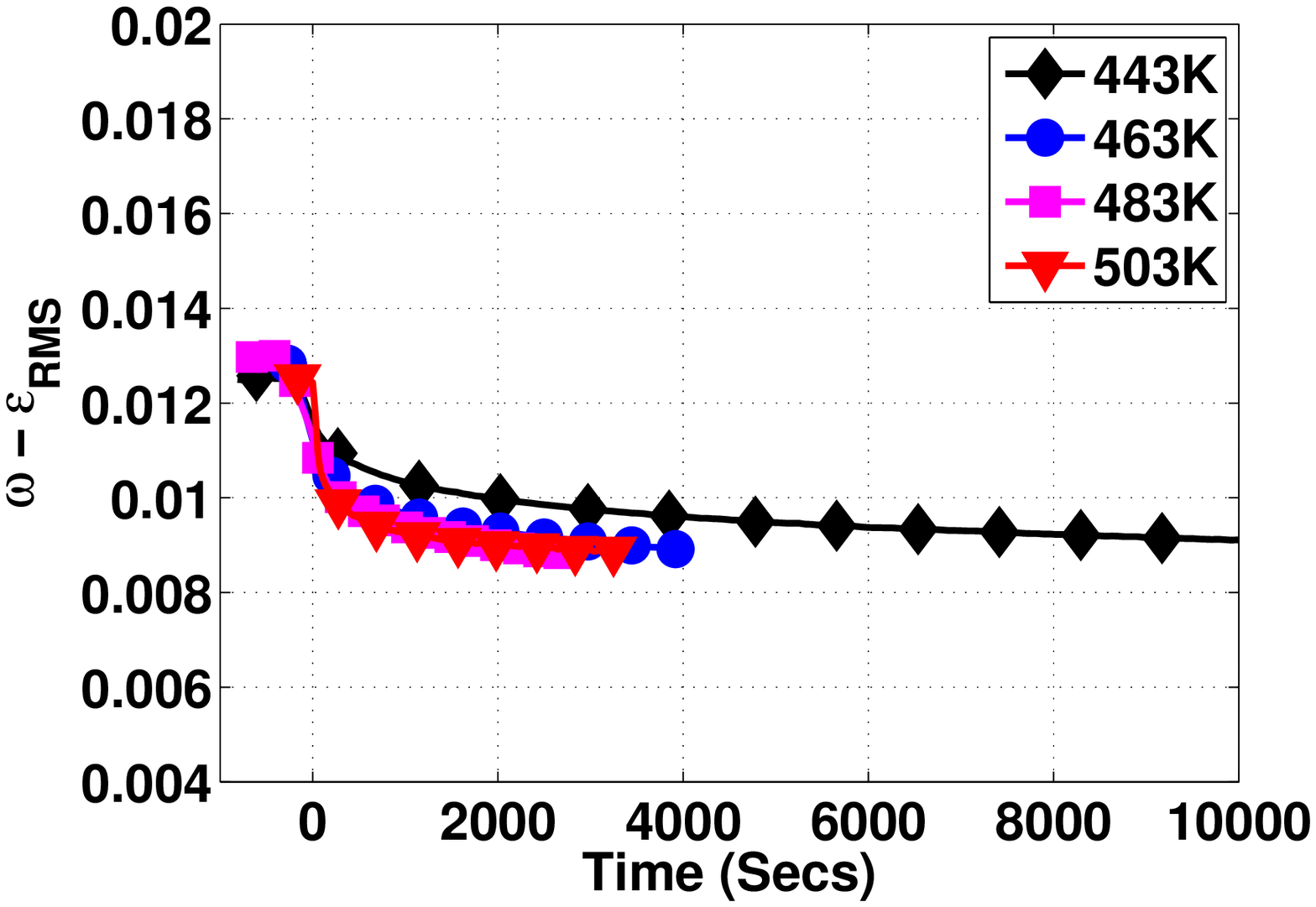}
\label{fig:rms1_d}
}

\end{tabular}
\end{center}
\caption{Comparison of the evolution of the RMS strains in both $\alpha$ and $\omega$ phase in the annealing of samples shocked to a peak stress of 8 GPa, (\protect\subref*{fig:rms1_a})-(\protect\subref*{fig:rms1_b}) and 10.5 GPa (\protect\subref*{fig:rms1_c})-(\protect\subref*{fig:rms1_d}). For visual clarity, markers are only shown every 10 or 15 data points. Note again that the RMS strain is an indicator of dislocation density.}  
\label{fig:rms1}
\end{figure*}

Figure \ref{fig:init_growth} shows the initial transformation rates of the 8 and 10.5 GPa shocked samples as a function of temperature. Transformations occuring during the temperature ramp-up is neglected in this calculation. The initial transformation rates increase with temperature for both samples as expected. Interestingly, a crossover trend is once again seen in the transformation rate as temperature changes. The sample shocked to 8 GPa starts with a higher initial transformation rate at the lowest annealing temperature (443K) but is eventually surpassed by the 10.5 GPa at higher temperatures. 

Figure \ref{fig:rms1} shows the time evolution of the RMS strain in both phases for material shocked to 10.5 GPa. Recall that the RMS strain $\varepsilon_{rms}$ serves as an indicator of the dislocation density. The initial $\varepsilon_{rms}$ values in the 4 samples shocked to its respective pressure displayed good repeatability, with $\varepsilon_{rms} \approx$ $0.01250\pm 2\times 10^{-5}$ for both 8 GPa and 10.5 GPa cases, indicating that the initial dislocation density in the $\omega$ phase was not shock pressure sensitive. In contrast, the RMS value for the $\alpha$ phase was $\sim 15\%$ higher in the 10.5 GPa shocked samples than in the 8 GPa shocked sample ($\approx0.017$ vs $\approx0.014$ respectively).

\begin{figure}[htb!]
\begin{center}
\includegraphics[scale=0.31,keepaspectratio=true]{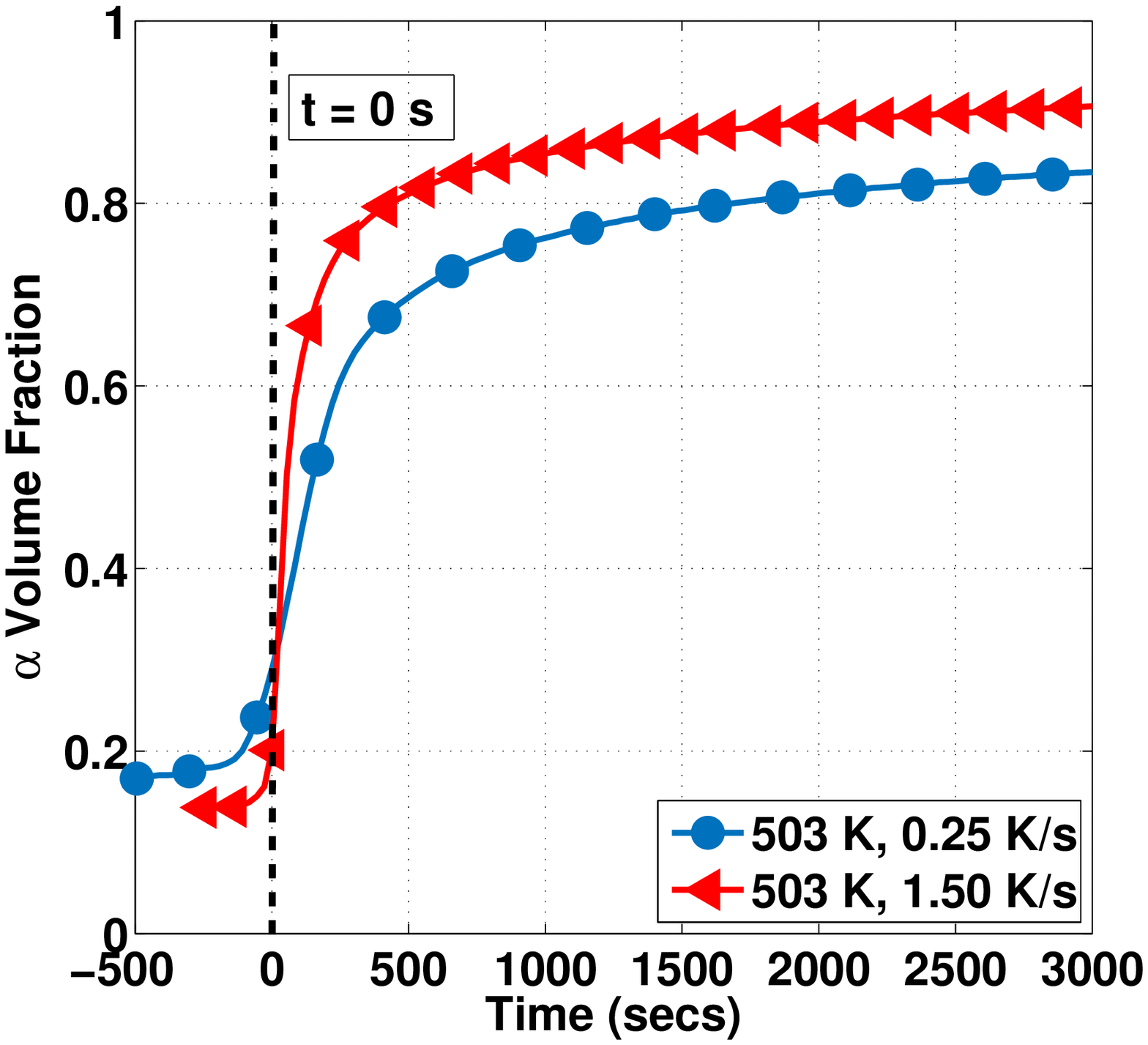}
\end{center}
\caption{Comparison of the evolution of the $\alpha$ fractions for different ramp rates to 503K. The slower ramp rate was at 0.25 K/s, the faster ramp rate was at 1.5 K/s.}
\label{fig:ramp}
\end{figure}

As described in Section \ref{sec:experimental_methods}, in order to explore the effects of thermal history on the microstructure evolution, a second set of samples was heated to 503K but at a slower heating rate (0.25K/s instead of 1.5K/s). Figure \ref{fig:ramp} shows a comparison of the evolution of $\alpha$ volume fraction for the 10.5 GPa samples. The 8 GPa sample displayed similar behavior and will be omitted for brevity. As seen in the figure,  the fast-ramp sample showed a higher initial transformation rate, transforming to a greater extent (higher $\alpha$ fraction), and saturating to the new metastable state faster than the slow-ramp samples. Furthermore, due to the slow heating-rate, a significant amount of transformation is observed before the sample reaches 503K (negative time in Figure \ref{fig:ramp} indicates heating). The differences in behavior clearly show that the microstructure evolution in the two-phase material is strongly dependent on the thermal history and the final temperature, further highlighting the non-equilibrium nature of the microstructural state. 

\subsection{Microstructure Characterization}

\begin{figure*}[p]
\centerline{
\begin{tabular}{>{\centering\arraybackslash} m{0.5in} >{\centering\arraybackslash} m{1.2in} >{\centering\arraybackslash} m{1.2in} >{\centering\arraybackslash} m{1.2in} >{\centering\arraybackslash} m{1.2in}}
  & Combined IPF Map & $\alpha$ IPF Map & $\omega$ IPF Map & Phase \& Interface\\
As-Shocked
	 & \subfloat[]{\includegraphics[width=3.2cm]{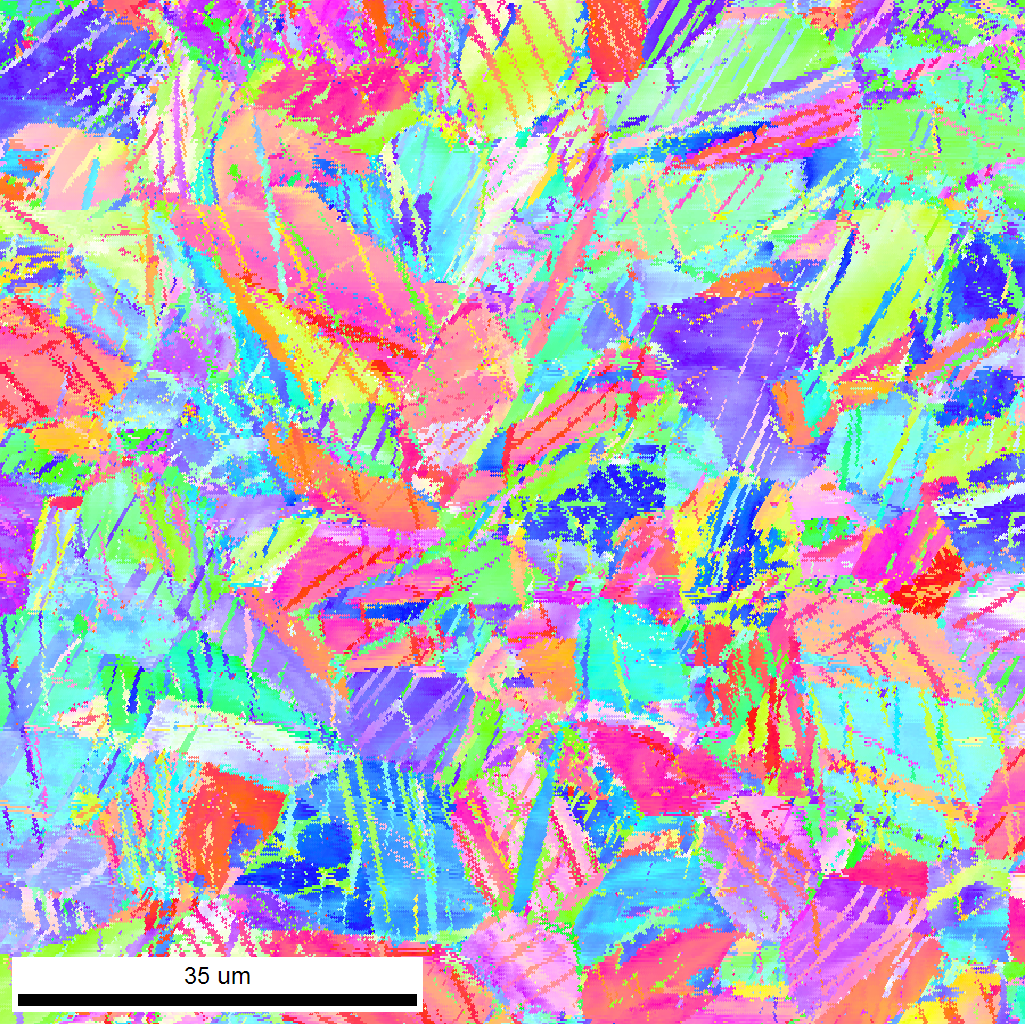} \label{fig:EBSDa}}
   & \subfloat[]{\includegraphics[width=3.2cm]{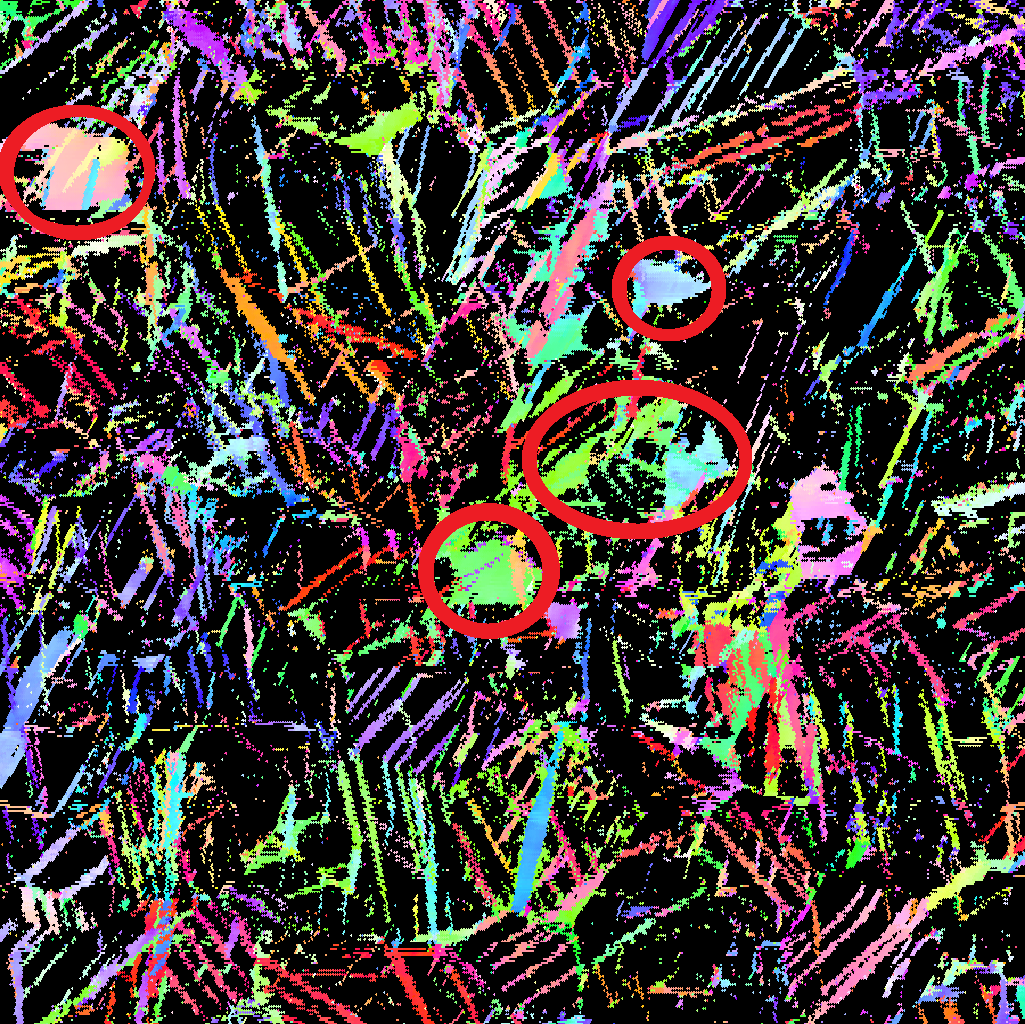} \label{fig:EBSDb}}
	 & \subfloat[]{\includegraphics[width=3.2cm]{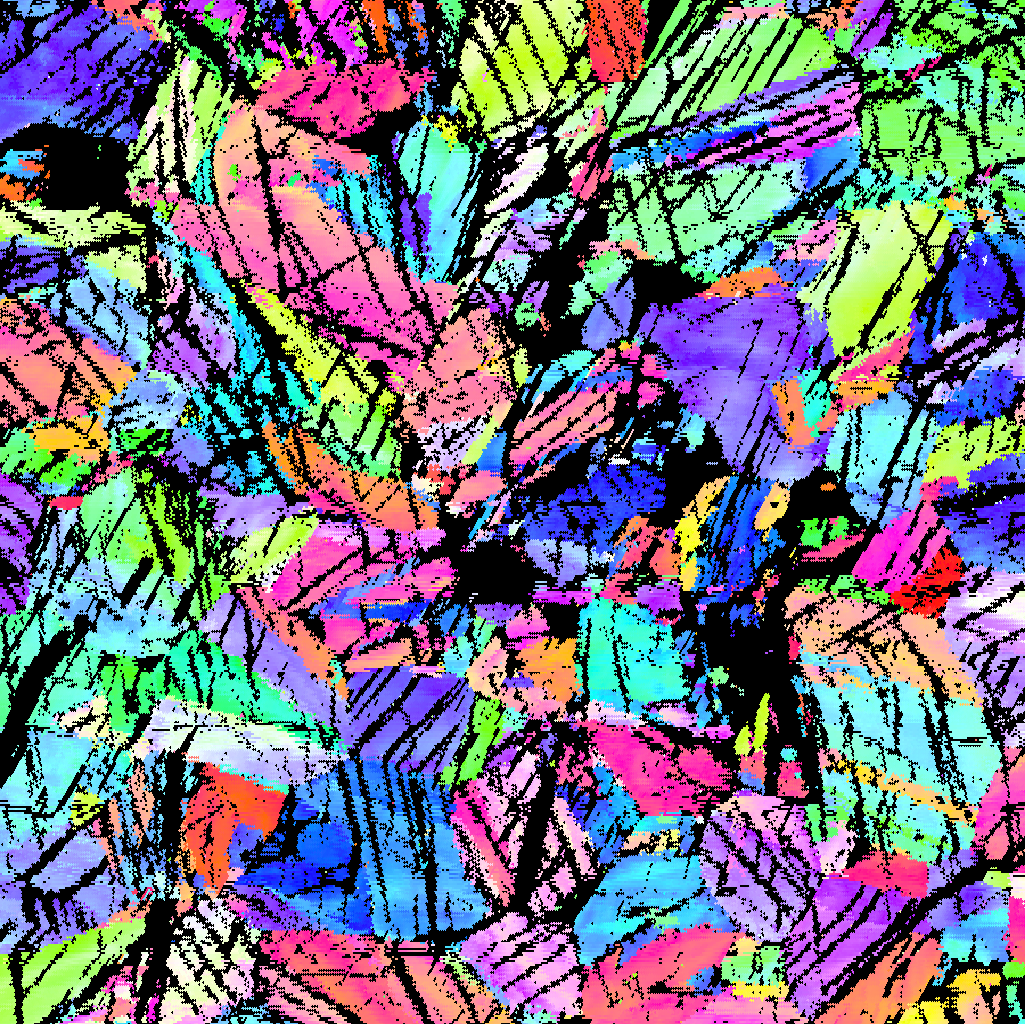} \label{fig:EBSDc}}
	 & \subfloat[]{\includegraphics[width=3.2cm]{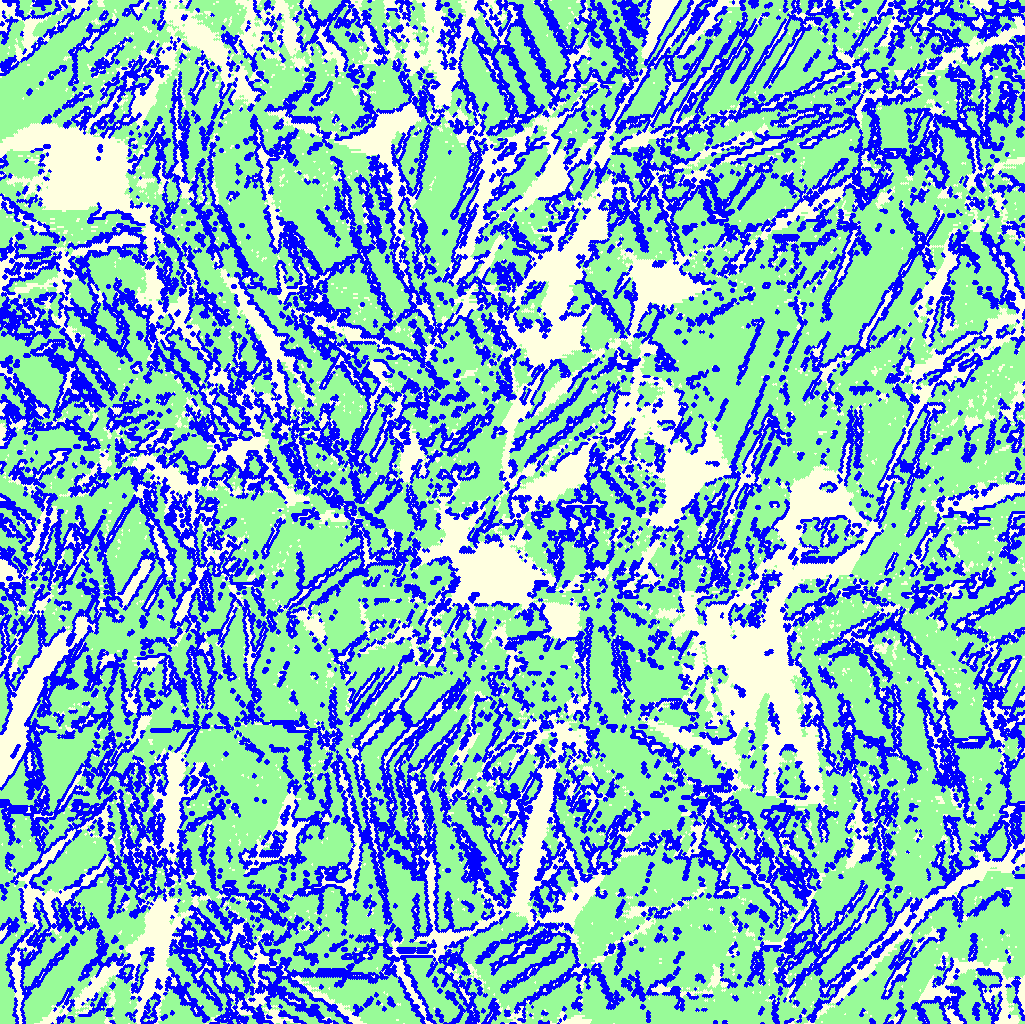} \label{fig:EBSDd}}\\
443K
	 & \subfloat[]{\includegraphics[width=3.2cm]{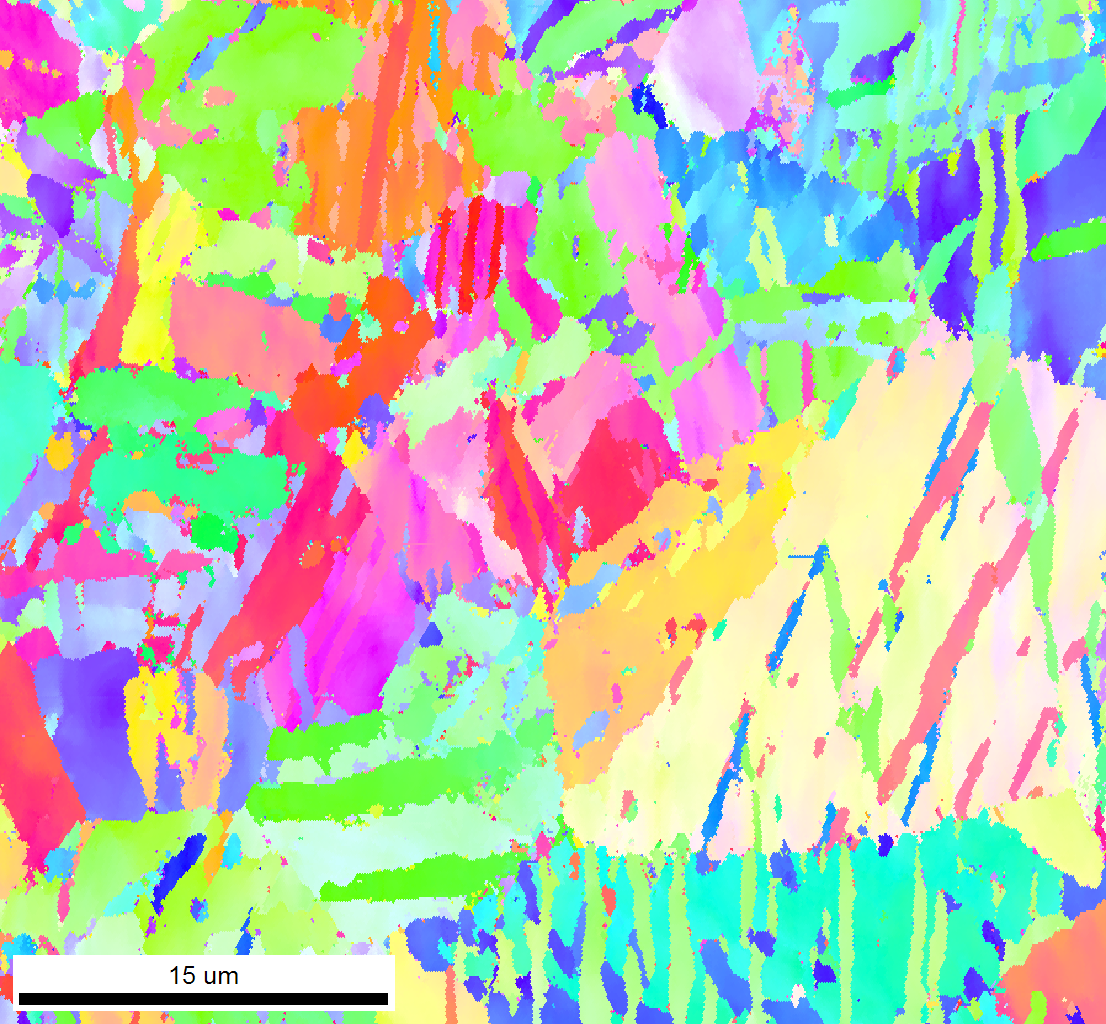} \label{fig:EBSDe}} 
   & \subfloat[]{\includegraphics[width=3.2cm]{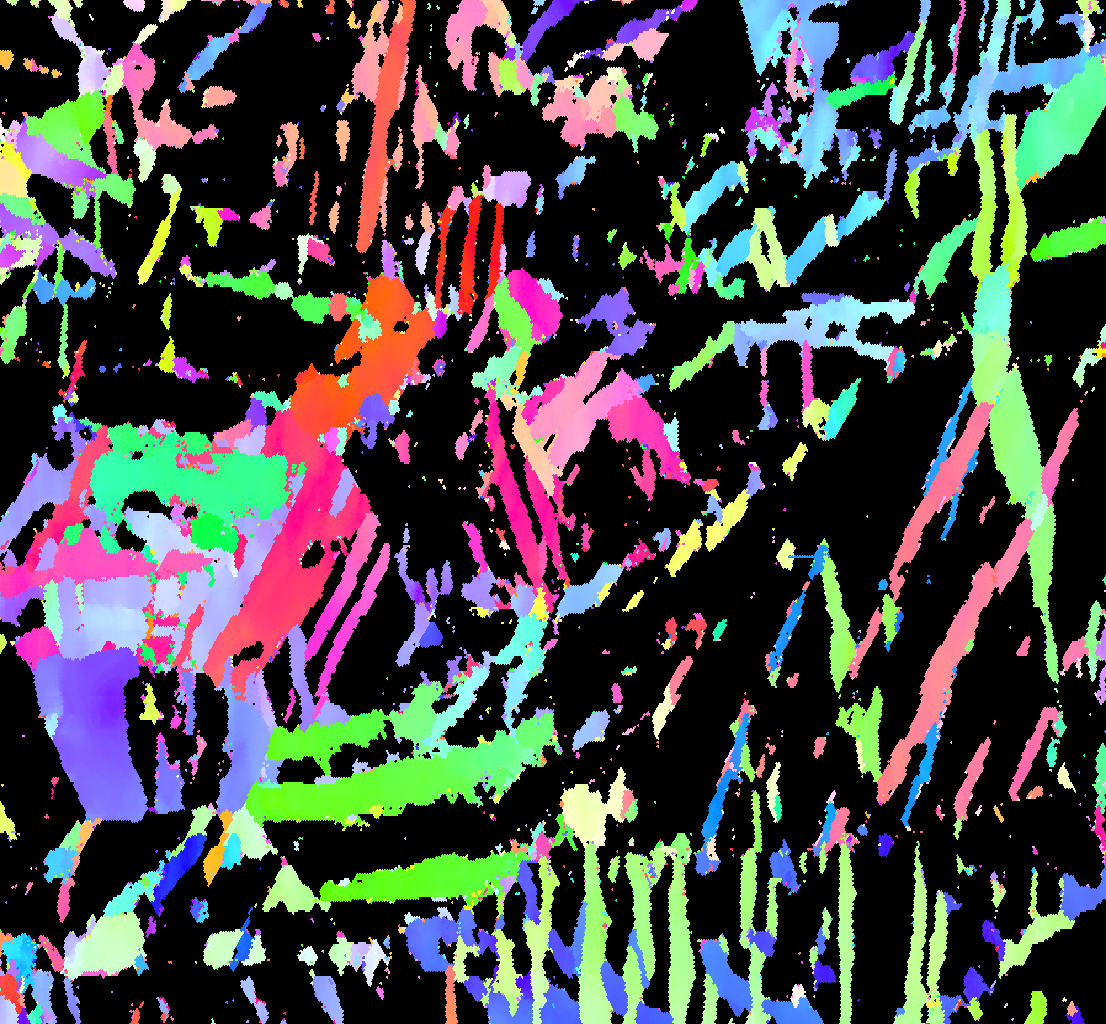} \label{fig:EBSDf}}
	 & \subfloat[]{\includegraphics[width=3.2cm]{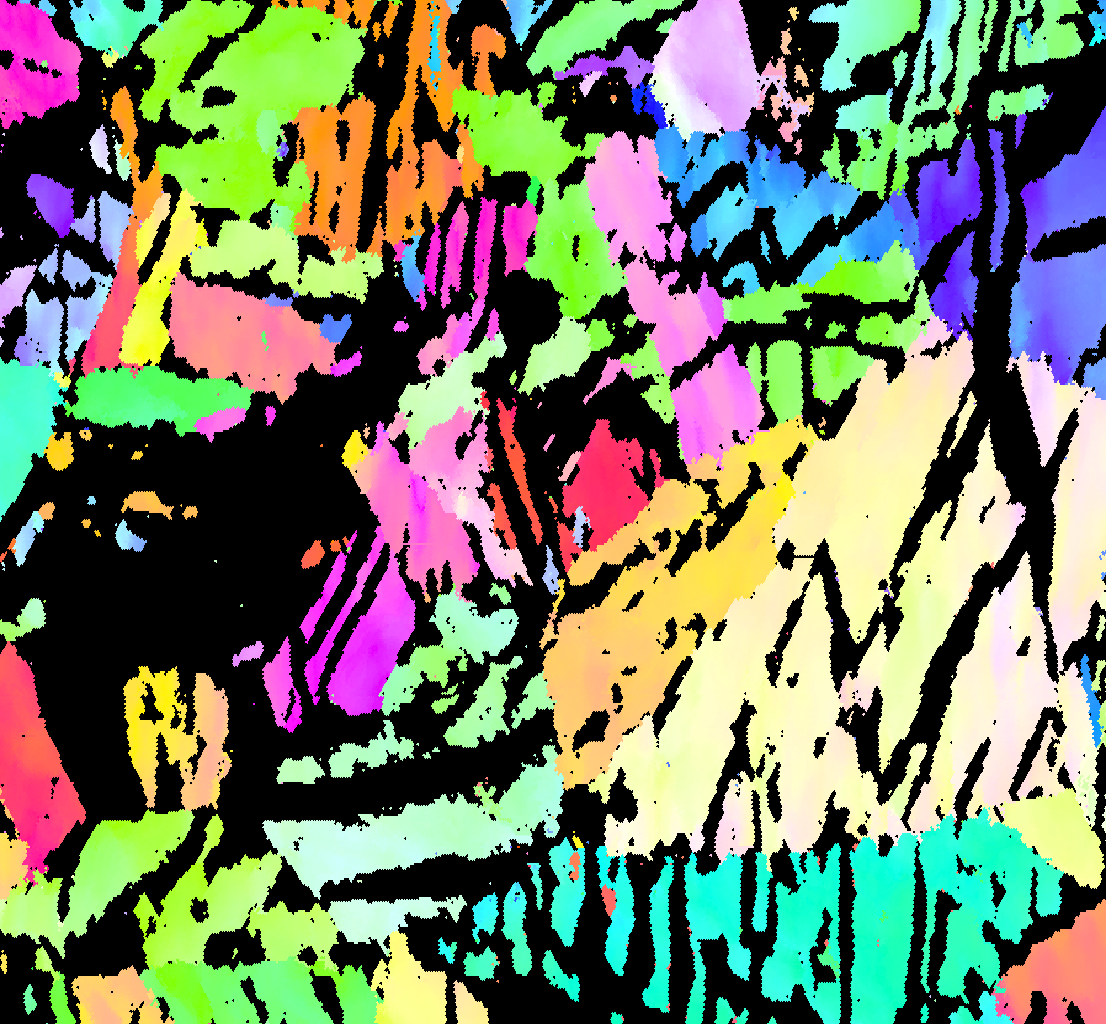} \label{fig:EBSDg}}
	 & \subfloat[]{\includegraphics[width=3.2cm]{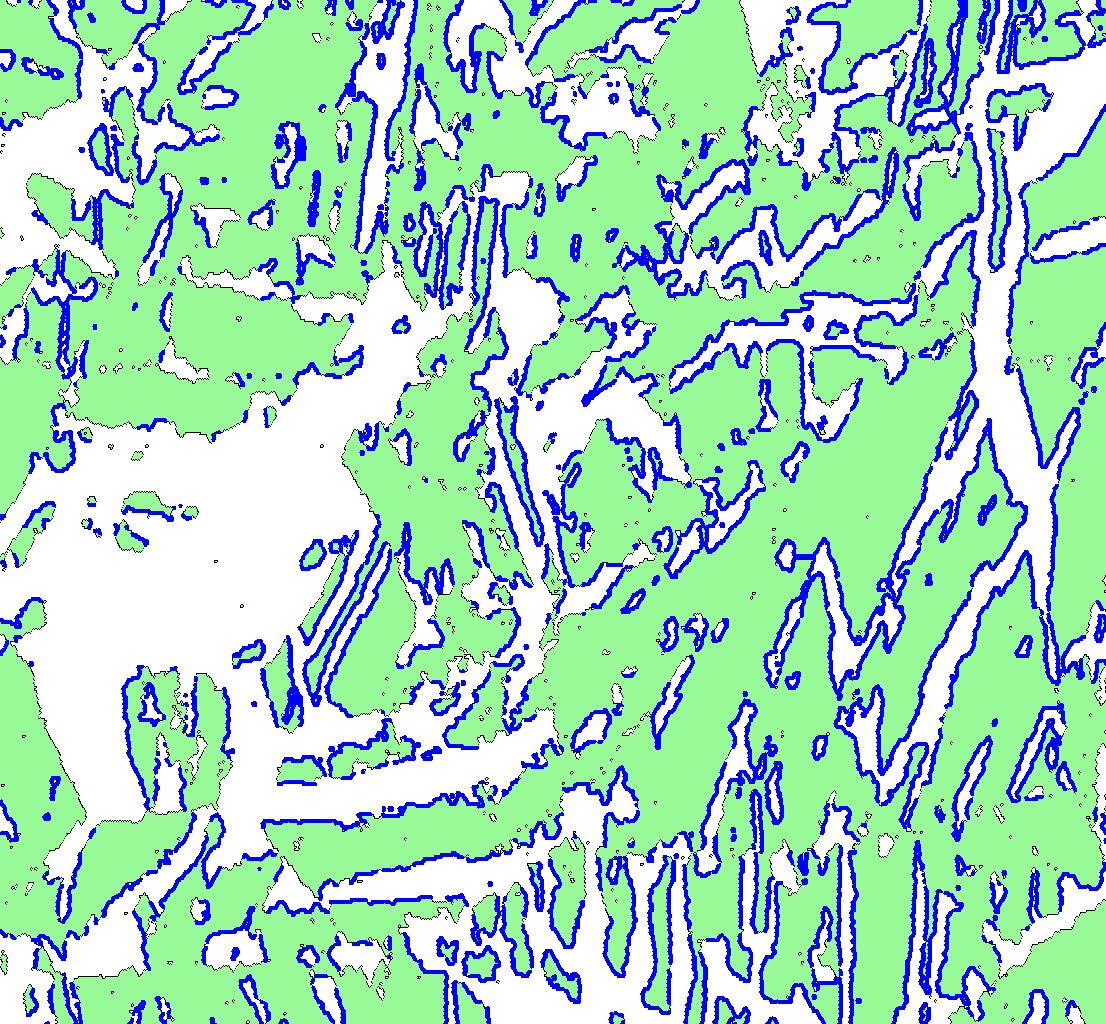} \label{fig:EBSDh}}\\
463K
	 & \subfloat[]{\includegraphics[width=3.2cm]{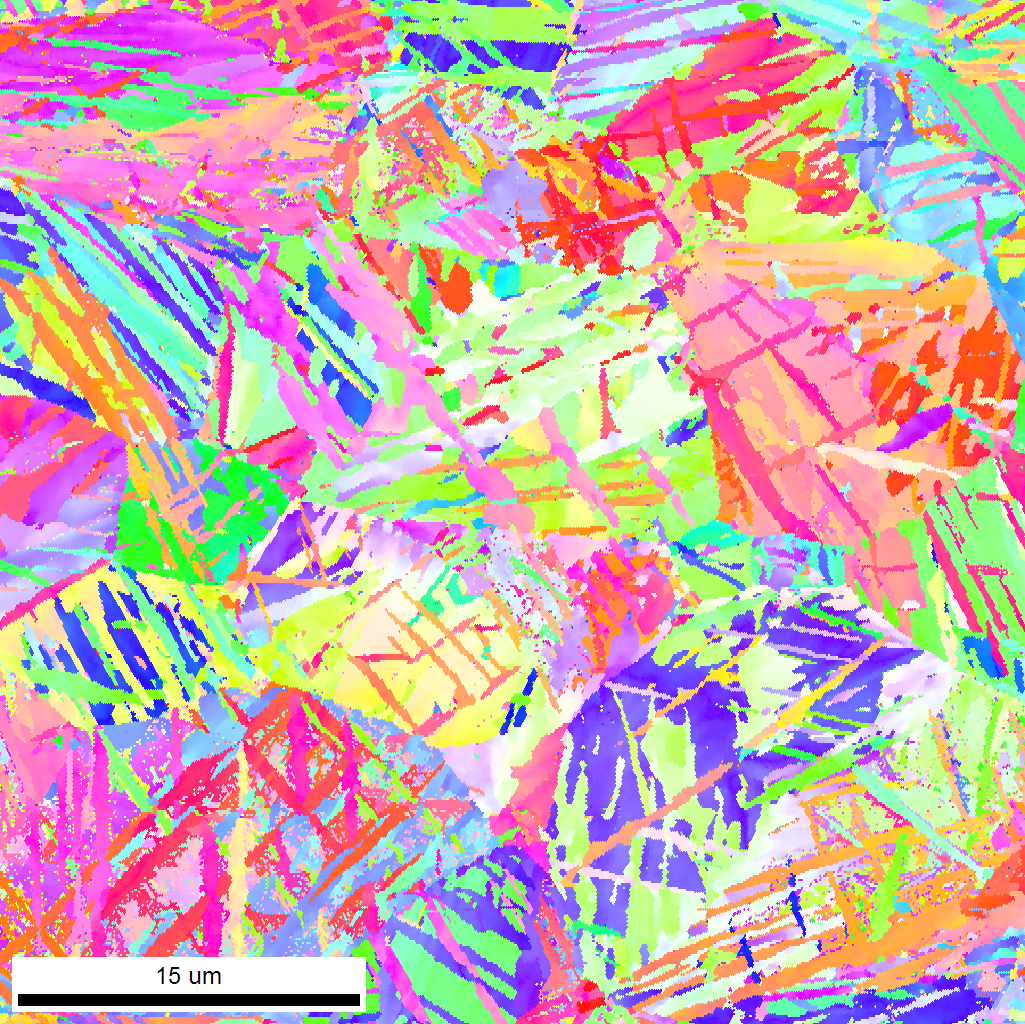} \label{fig:EBSDi}} 
   & \subfloat[]{\includegraphics[width=3.2cm]{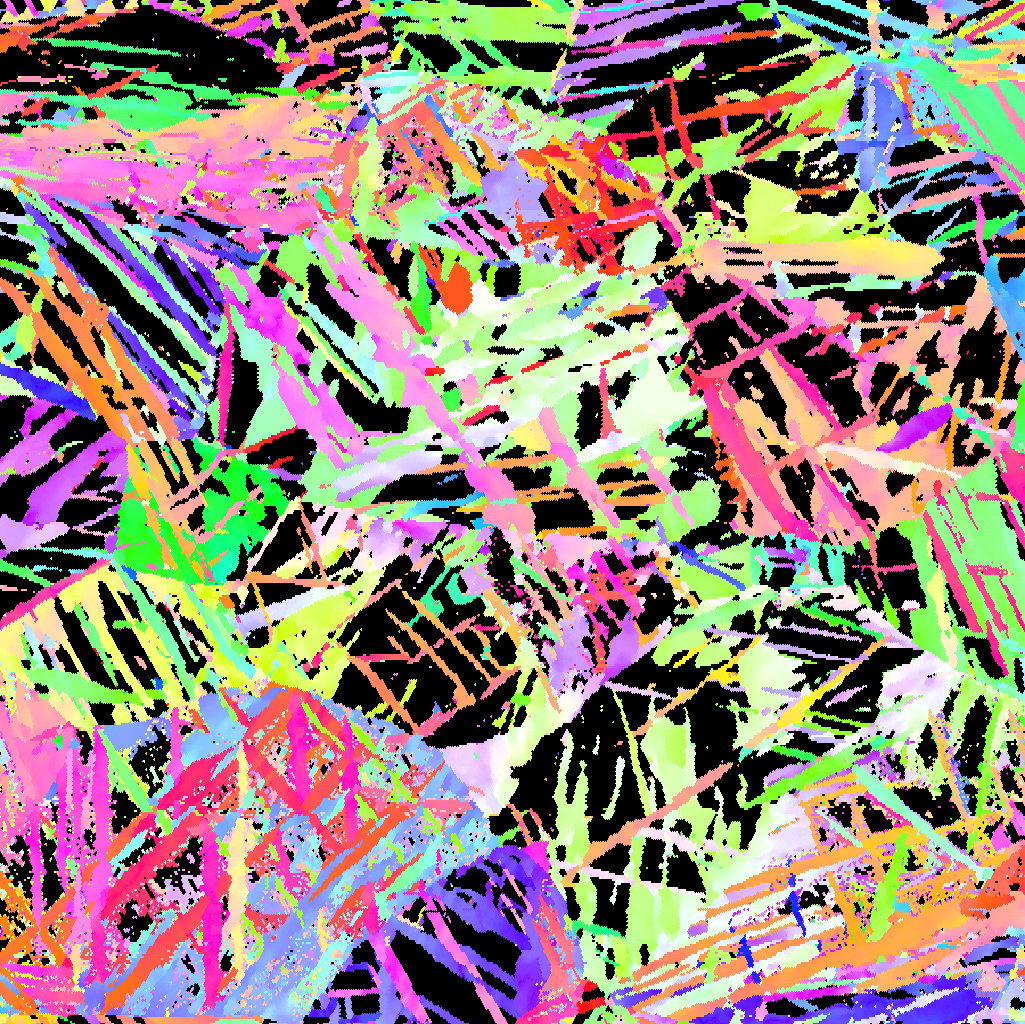} \label{fig:EBSDj}}
	 & \subfloat[]{\includegraphics[width=3.2cm]{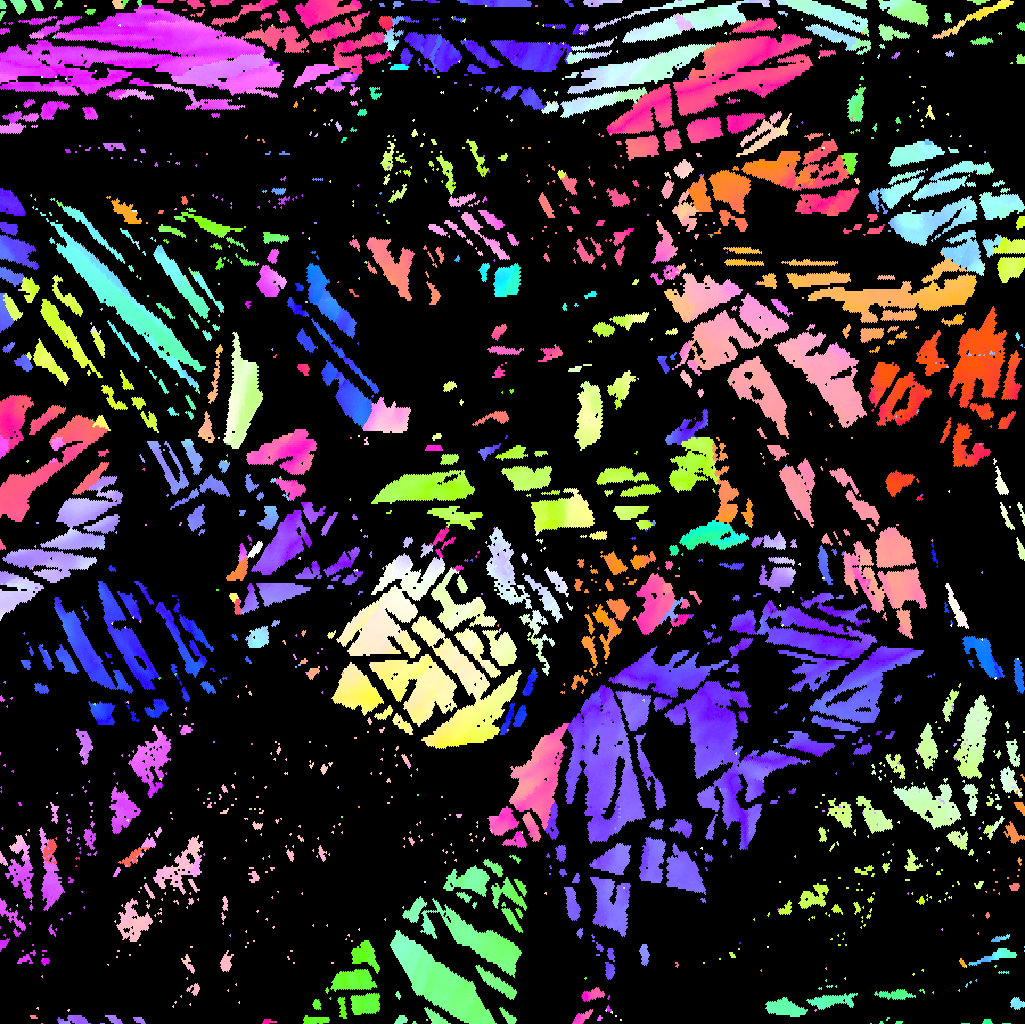} \label{fig:EBSDk}}
	 & \subfloat[]{\includegraphics[width=3.2cm]{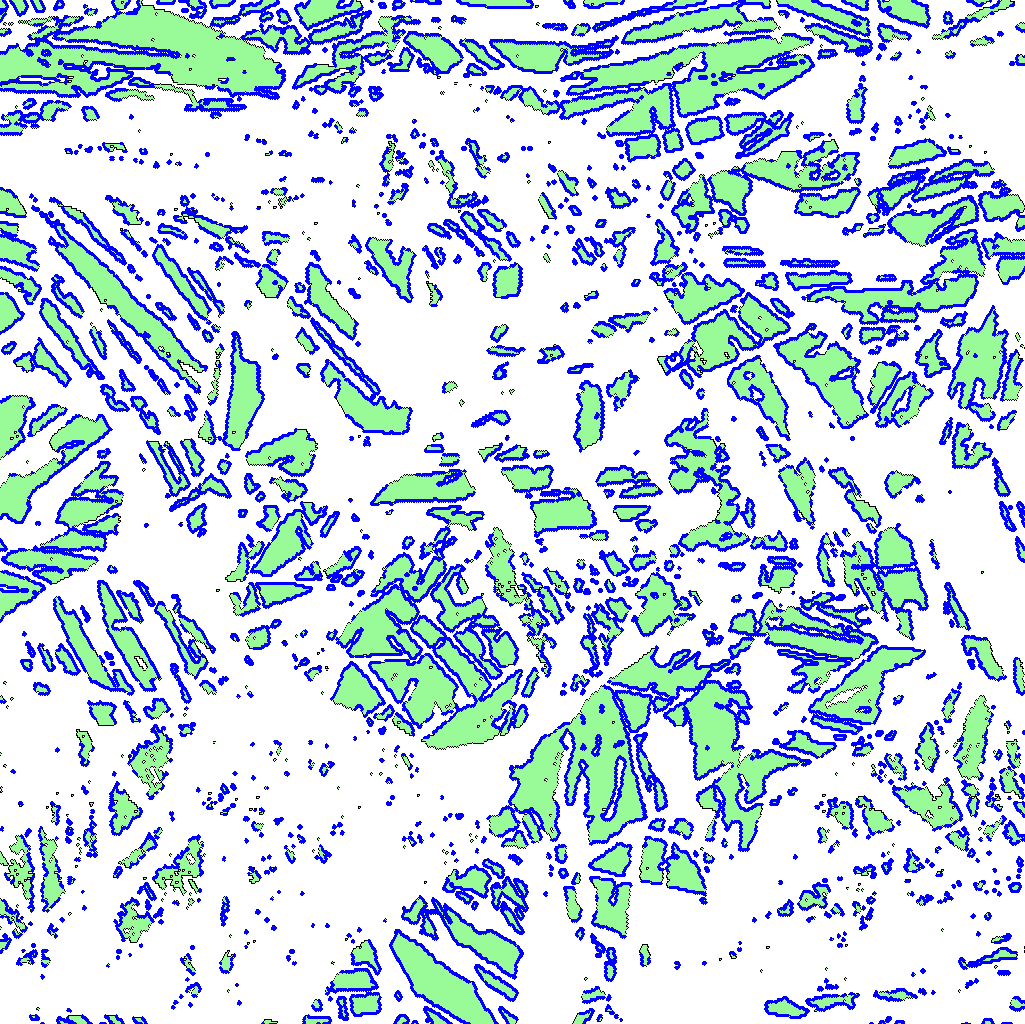} \label{fig:EBSDl}}\\
Color legend
	 & \subfloat[]{\includegraphics[width=3.2cm]{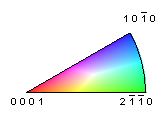} \label{fig:color_legend}}
\end{tabular}
}

\caption{EBSD images obtained from OIM Software, with the shock direction normal to the analyzed surface. (a)-(d) correspond to the non-annealed sample, (e)-(h) correspond to sample annealed at 443K, and (i)-(l) correspond to sample annealed at 463K. The Phase and Interface figures (d), (h), and (l) are color coded as such: $\alpha$-white, $\omega$-light green. The blue lines mark interfaces with the $(0\;0\;0\;1)_\alpha\parallel(1\;0\;\overline{1}\;1)_\omega$ and $[1\;0\;\overline{1}\;0]_\alpha\parallel[1\;1\;\overline{2}\;\overline{3}]_\omega$ orientation relationship.}
\label{fig:EBSD}
\end{figure*}

Figure \ref{fig:EBSD} shows orientation and phase maps of $90\times90 ~\mu$m regions in the (8 GPa) as-shocked and annealed samples. For the annealed specimens, samples were heated at 443K and 463K for a total duration of 21,336 and 19,145 seconds respectively. As seen in Figure \ref{fig:EBSD}\subref*{fig:EBSDb}, the as-shocked $\alpha$ phase grains are lamellar with grains that are reminiscent of deformation twins or fine martensite. The $\omega$ grains (Figure \ref{fig:EBSD}\subref*{fig:EBSDc}) make a relatively equiaxed matrix in which the $\alpha$ lathes are embedded. The circled $\alpha$ grains in the as-shocked $\alpha$ phase orientation map (Figure \ref{fig:EBSD}\subref*{fig:EBSDb}) are hypothesized to be the un-transformed remnants of original pre-shock microstructure. This hypothesis is based on the morphological differences (equiaxed vs lath) and the presence of $\lbrace 1 0 \overline{1} 1\rbrace$ compression twins. Figure \ref{fig:EBSD}\subref*{fig:EBSDd} confirms the lath like $\alpha$ grains exhibit an $(0\;0\;0\;1)_\alpha\parallel(1\;0\;\overline{1}\;1)_\omega$ and $[1\;0\;\overline{1}\;0]_\alpha\parallel[1\;1\;\overline{2}\;\overline{3}]_\omega$ orientation relationship with the equiaxed parent $\omega$, as reported in previous studies on shocked Zr \cite{song1995microscopic}. That this relationship is not observed in the circled grains in Figure \ref{fig:EBSD}\subref*{fig:EBSDb} supports that they are untransformed remnants of the original structure and provides further evidence for the hypothesis that, during the shock, nearly all of the initial $\alpha$ microstructure is transformed to $\omega$ and that majority of the observed $\alpha$ grains post-shock are children of the $\omega$ structure \cite{PhysRevB.89.220101}.

Figures \ref{fig:EBSD}\subref*{fig:EBSDe}, \ref{fig:EBSD}\subref*{fig:EBSDf}, and \ref{fig:EBSD}\subref*{fig:EBSDg} show the microstructure after partial annealing at 443K. As displayed, the $\alpha$ laths have coarsened considerably relative to the as-shocked microstructure. The area fraction of $\alpha$ from EBSD analysis in the partially annealed structure is 0.41 as opposed to 0.32 in the as-shocked structure. As measured orthogonal to the long axis of the $\alpha$ grain, the average lath thickness of the 443K anneal material is $0.79\pm0.30~\mu$m vs $0.51\pm0.21~\mu$m in the as-shocked structure. This analysis disregards the 3D orientation of the laths with respect to the cutting plane, and is not meant to represent a true stereological width of the $\alpha$ grains, rather it is meant as a simple comparative metric between the as-shocked and annealed microstructures (assuming that the crystallographic texture of the $\alpha$ phase is fairly uniform and that there is no preferential growth of different $\alpha$ variants at different temperatures).  In contrast, the material heated at 463K (Figures \ref{fig:EBSD}\subref*{fig:EBSDi}, \ref{fig:EBSD}\subref*{fig:EBSDj}, and \ref{fig:EBSD}\subref*{fig:EBSDk}) has a microstructure composed of both coarsened and very fine laths. In fact, the average lath thickness is nearly identical to the as-shocked material ($0.50\pm0.16\mu$m) despite transforming to 0.74 area-fraction $\alpha$. This implies that both growth of the existing $\alpha$ grains and nucleation of new $\alpha$ laths is occurring at 463K, but that no or very limited nucleation of new $\alpha$ is taking place at 443K and that the transformation is progressing completely by the growth of existing $\alpha$ grains. Figures  \ref{fig:EBSD}\subref*{fig:EBSDi} and  \ref{fig:EBSD}\subref*{fig:EBSDl}, show that the expected $\alpha/\omega$ orientation relationship is maintained during the nucleation and growth process. 

\section{Discussion}
\label{sec:discussion}

From purely thermodynamic considerations, the $\alpha$ and $\omega$ phases in a pure Zr material should not co-exist  at standard temperature and pressure (STP) conditions. Despite the $\alpha$ phase being the equilibrium structure at STP, it is well established that post shock microstructures with $\omega$ fractions $\ge0.8$ are metastable at STP conditions for years following a shock induced phase transformation \cite{Cerreta20137712}. 

The working hypothesis in developing the above experimental study was that the complex dislocation state in the $\omega$ phase arrests the reverse transformation, and that the reduction of dislocation density in the $\omega$ phase with sufficient heating allows the reverse transformation to advance. These arresting dislocation structures are postulated to have been mobile in the pre-shock microstructure but are likely sessile in the $\omega$ phase. Several critical questions directly follow from this working hypothesis. These include: i) What is the asymptotic behavior when the material is isothermally heated? ii) How sensitive is the material response to thermal history and heating rate? and iii) What is the balance between nucleation of new $\alpha$ grains and growth of existing $\alpha$ grains? The heating experiments described above were developed to begin addressing these questions and particularly to provide the required data for the development of microstructure sensitive kinetic models of the transformation. The X-ray and EBSD analysis results indicate a more complex transformation behavior than suggested by earlier experiments and providing critical insight into the microstructure evolution during heating. 

Several distinct martensitic or shear/shuffle transformation mechanisms have been proposed for the forward transformation \cite{Rabinkin, jyoti2008, usikov1973orientation, SarathKumarMenon1982717, song1995microscopic, Gupta19851167}. Focusing on Figure \ref{fig:alpha_evolution_1}\subref*{fig:alpha_evolution_2_a} and \ref{fig:alpha_evolution_1}\subref*{fig:alpha_evolution_2_b}, attention is immediately drawn to the temperature dependence of the initial $\alpha$ volume fraction growth rates and the extent of reverse transformation from $\omega \to \alpha$. In general, a strong temperature dependence of transformation rate between two phases without change in chemical composition would indicate a massive or short range-diffusional type mechanism; where the rate would be directly related to the difference in phase fraction. However, comparing the heating of samples in Figures \ref{fig:alpha_evolution_1}\subref*{fig:alpha_evolution_2_c} and \ref{fig:alpha_evolution_1}\subref*{fig:alpha_evolution_2_d} reveals a cross over of $\alpha$ volume fractions between the two different shock loaded samples. Despite having similar $\alpha$ phase concentrations at the crossing point, the transformation rates still differ, indicating that the driving force is not solely based on the concentration difference between the phases, but also on the microstructure of the material. Keeping this in mind, the observation of initially high transformation rates followed by a plateau to a metastable state with significantly retained $\omega$ also indicates a limiting of the reverse transformation at STP conditions due to the microstructural state of the $\omega$ phase. This idea, taken together with the low self-diffusivity of Zr at a low homologous temperature ($10^{-23}m^2/s$ at 800K)\cite{Horváth1984206}, strongly points to a shear mechanism for the the reverse transformation as well. One can then deduce that the temperature dependency is sourced from dislocation of defect pinning within the $\omega$ phase, although the exact dislocation reactions and mechanisms, e.g. local dislocation annihilation, cutting or unlocking, which facilitate the transformation are still unknown.

 A dislocation mediated shear mechanism is consistent with very recent atomistic modeling by Zong et al., informed by preliminary analysis of the data presented here \cite{zong2014}. In that work the evolution of $\omega$ fraction is fit to a modified Kohlrausch-Williams-Watts equation  \cite{PhysRevB.37.3716}or stretched exponential of the form $\eta(t)\propto t^{-\alpha} \cdot \exp{\left(- \left(\tfrac{t}{\tau(T)} \right)^\beta \right)}$, there $\eta$ is the $\omega$ fraction, $t^{-\alpha}$ is an pre-factor that accounts for non-thermally activated events, $T$ is the absolute temperature, and $\tau$ is the relaxation time or time for $\approx$37\% of the transformation to take place. The produced fits show that the $\alpha$ was close to zero indicating that that the thermal contribution to the kinetics was small and the exponent $\beta\approx 0.5$ indicating that the kinetics are analogous to other ``glassy" or ``frustrated" systems and not consistent with Johnson-Mehl-Avrami-Kolmogorov nucleation and growth \cite{PhysRevB.87.094109,PhysRevB.63.174101}. The predicted kinetics of the $\omega \to \alpha$ transformation in Ti with a simulated shocked microstructure via molecular dynamics was qualitatively consistent with the experimental results  for Zr. However, in that work  Zong suggests that nucleation is energetically preferable to growth by interface migration, which is inconsistent with our EBSD results which show significant growth at 443K without observable nucleation.

Recall that the RMS strain values qualitatively track the dislocation density in each phase. Upon heating, $\varepsilon_{rms}$ decreases quickly in both samples and both phases as evidenced in Figure \ref{fig:rms1}. The evolution of dislocation densities in both phases are distinct -- the rate at which $\varepsilon_{rms}$ decreases in the $\alpha$ phase is strongly temperature dependent, whereas the decrease of $\varepsilon_{rms}$ in the $\omega$ phase showed minimal temperature dependence, with the exception of the heating at 443K. This points towards distinct mechanisms behind the evolution of the microstructure in the individual phases. 

\begin{figure*}[htb!]
\begin{center}

\begin{tabular}{cc}

\subfloat[$\varepsilon_{rms}$ in $\alpha$ phase (8GPa)]{
\includegraphics[scale=0.31,keepaspectratio=true]{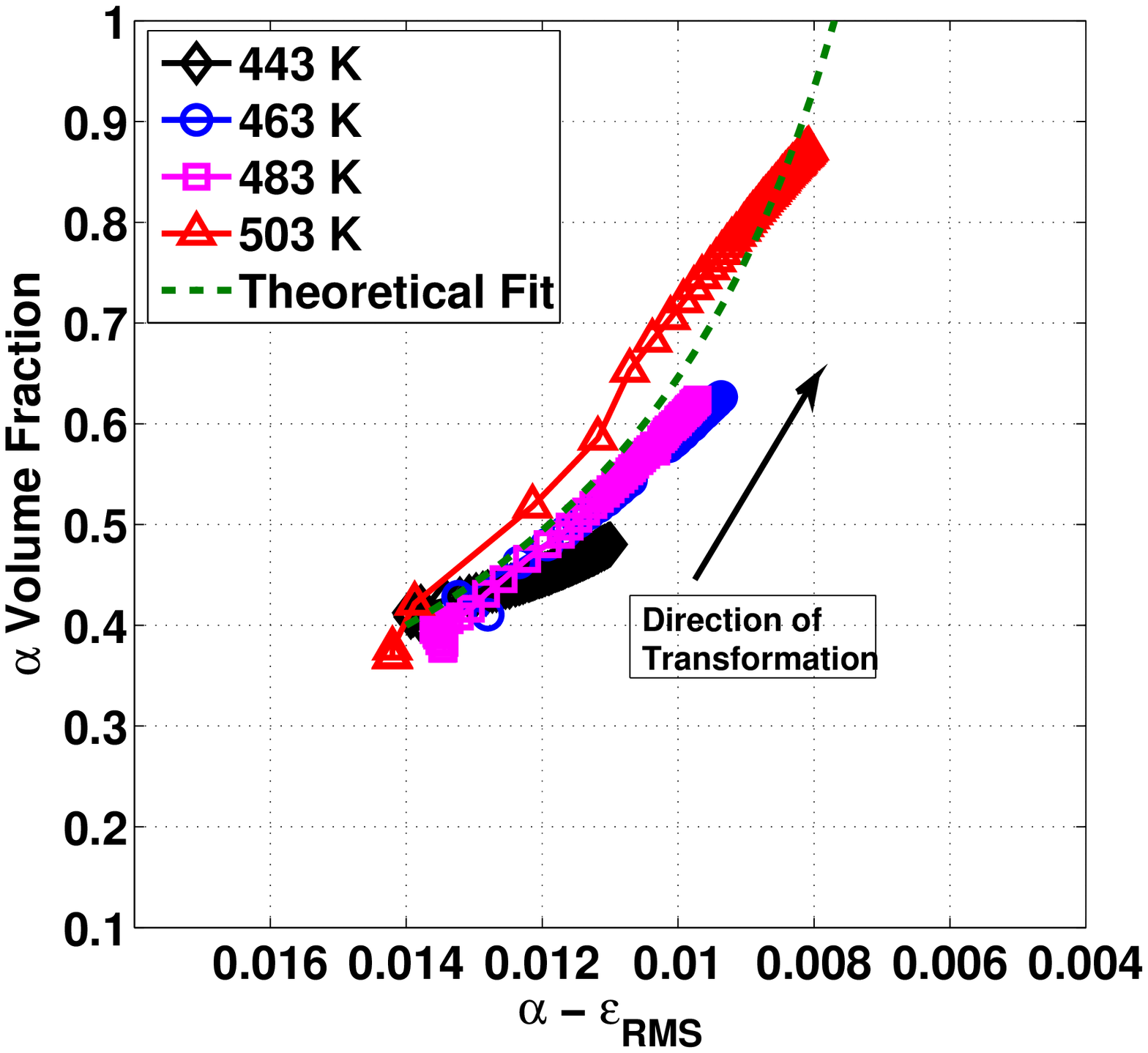}
\label{fig:rms2_c}
} 

\subfloat[$\varepsilon_{rms}$ in $\omega$ phase (8GPa)]{
\includegraphics[scale=0.31,keepaspectratio=true]{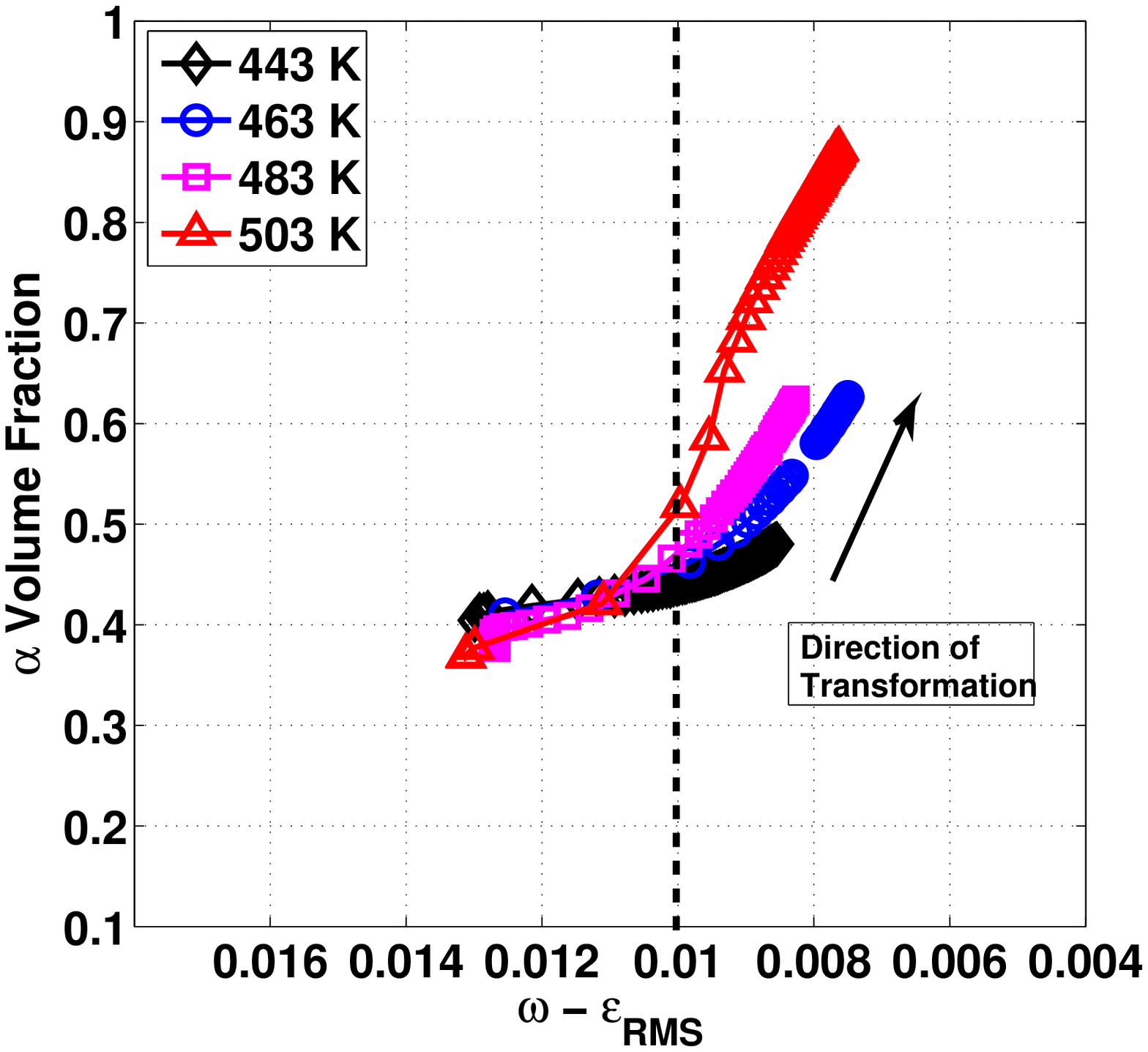}
\label{fig:rms2_d}
} \cr

\subfloat[$\varepsilon_{rms}$ in $\alpha$ phase (10.5GPa)]{
\includegraphics[scale=0.31,keepaspectratio=true]{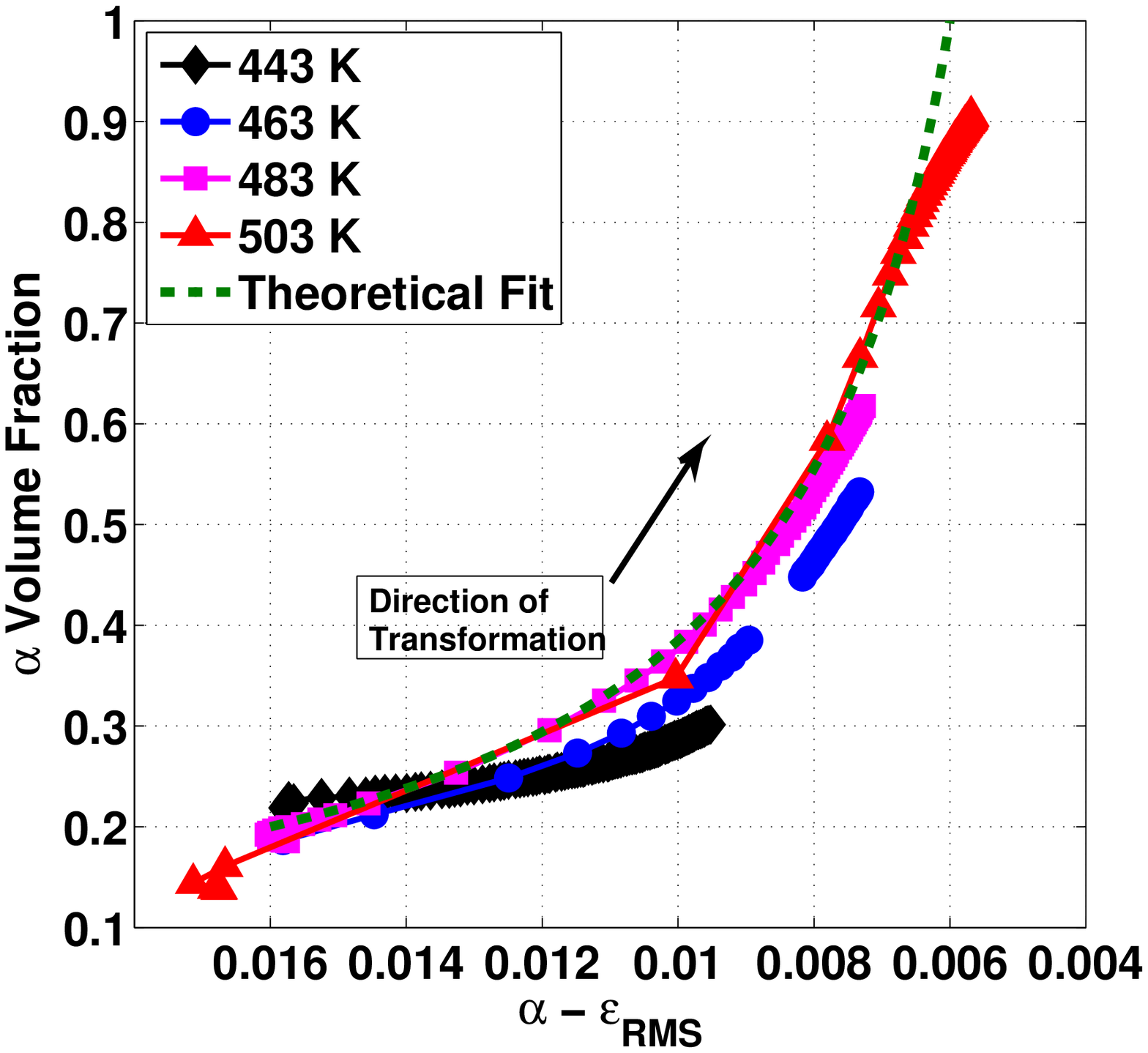}
\label{fig:rms2_a}
}

\subfloat[$\varepsilon_{rms}$ in $\omega$ phase (10.5GPa)]{
\includegraphics[scale=0.31,keepaspectratio=true]{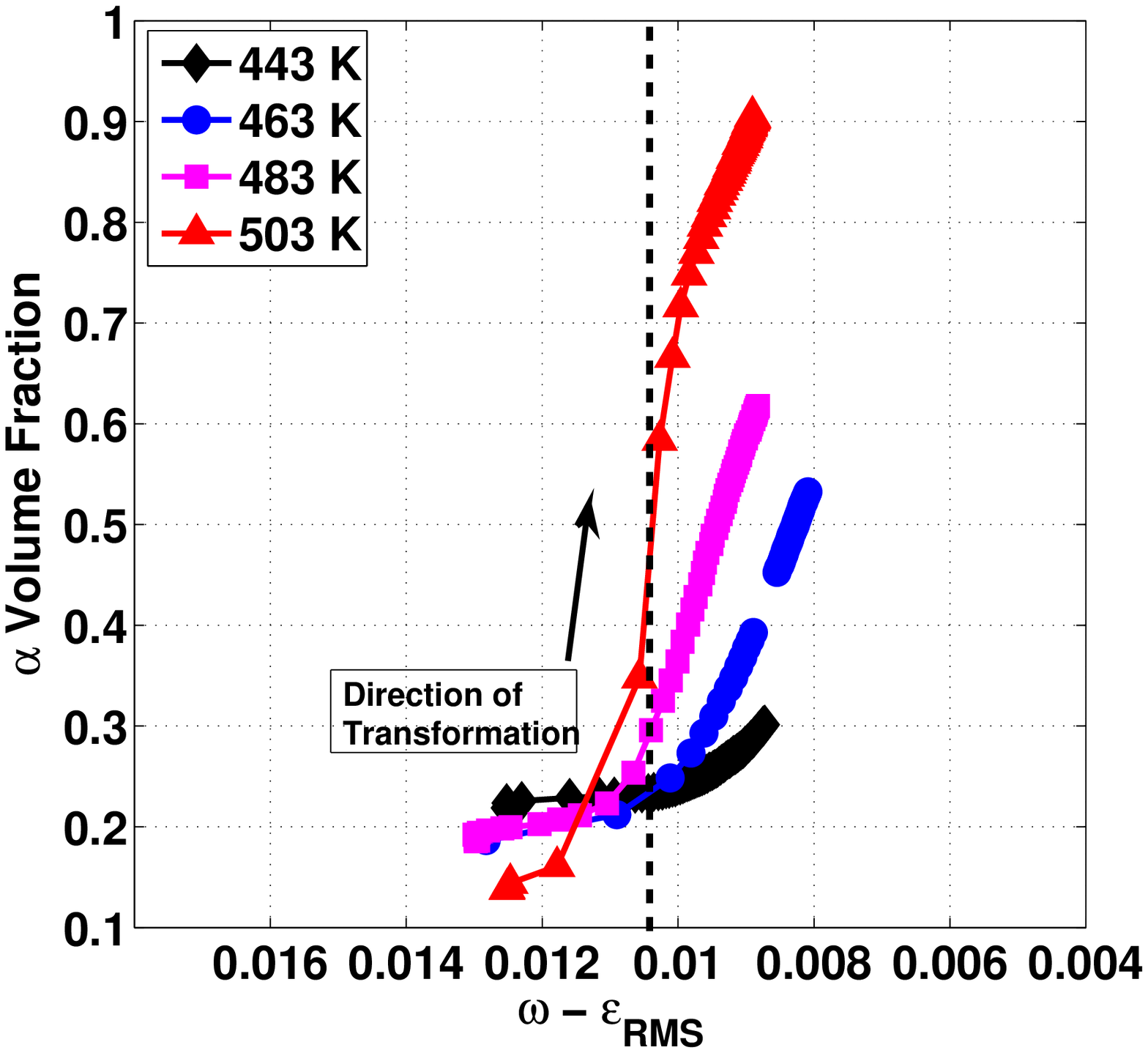}
\label{fig:rms2_b}
}

\end{tabular}
\end{center}
\caption{Comparison of the evolution of RMS strain relative to the $\alpha$ phase volume fraction evolution in the annealing of samples shocked to a peak stress of 10.5 GPa in the (\protect\subref*{fig:rms2_c})--(\protect\subref*{fig:rms2_a}) $\alpha$ phase and (\protect\subref*{fig:rms2_d})--(\protect\subref*{fig:rms2_b}) $\omega$ phase. The theoretical fit in (\protect\subref*{fig:rms2_c}) and (\protect\subref*{fig:rms2_a}) is obtained using a weighted average method over the original and incipient $\alpha$ phase.} 
\label{fig:rms2}
\end{figure*}

For the $\alpha$ phase, we suggest that the reduction of the dislocation density is due to the creation of incipient $\alpha$ phase material with a relatively lower dislocation density, resulting in a reduction of the average dislocation density in the $\alpha$ phase. If we assume that the $\alpha$ phase present following the shock ($\nu_{\alpha_0}$) has a fixed RMS strain ($\varepsilon_{\alpha_0}$) and that the incipient $\alpha$ ($\nu_{\alpha_i}$)  also has a fixed lower RMS strain ($\varepsilon_{\alpha_i}$), then the average RMS strain could be approximated by a weighted average of the two components, given by:
\begin{equation}
\varepsilon_{rms,\alpha} = \frac{(\varepsilon_{\alpha_0}\nu_{\alpha_0} + \varepsilon_{\alpha_i}\nu_{\alpha_i})}{\nu_{\alpha_i} + \nu_{\alpha_0}}
\label{eq:weighted_average}
\end{equation}
Using an appropriate initial volume fraction of $\alpha$ for each material -- 0.4 and 0.2 in the material shocked to 8 and 10.5 GPa respectively, and likewise assigning RMS strain values of  0.014 and 0.016 to the as-shocked $\alpha$ and 0.0035 to the incipient $\alpha$, an RMS strain averaged over the $\alpha$ phase is calculated and plotted in Figure \ref{fig:rms2}.  The simple calculation fits the data observed during heating of the material shocked to 10.5 GPa very well, and is reasonable for the 8 GPa material. It is worth re-stating at this point that the curve for annealing at 443K remains an exception, reinforcing the notion of a different mechanism (growth vs nucleation) for the reverse transformation depending on a temperature threshold. More on this mechanism difference will be presented in the discussion of EBSD results.

Figures \ref{fig:rms2}\subref*{fig:rms2_d} and \ref{fig:rms2}\subref*{fig:rms2_b} show the evolution of the $\alpha$ volume fraction as a function of the $\omega$ RMS strain. The arrows indicate the direction of evolution on these plots. Notice that the dislocation density in the $\omega$ phase decreases considerably before the reverse transformation begins in earnest, indicated by the significant increase in $\alpha$. Indeed, it appears that the defect level in the $\omega$ microstructure must be reduced to a critical value, corresponding to an RMS strain of approximately 0.01, for the initiation of the reverse transformation.  The mechanism behind the removal of these arresting dislocations is still unknown, but it can be further inferred that the energy required to remove the arresting dislocations in the $\omega$ phase contributes significantly to the nucleation energy barrier. 

In comparing the microstructure states for different shock loadings, the initial $\alpha$ dislocation density was about 15\% higher in the 10.5 GPa sample than in the 8 GPa sample.(Figures \ref{fig:rms1}\subref*{fig:rms1_a} and \ref{fig:rms1}\subref*{fig:rms1_c}). During the initial stages of the shock prior to the transformation from $\alpha$ to $\omega$, the initial $\alpha$ microstructure undergoes substantial plastic deformation and accumulates a significant dislocation density. On transforming to $\omega$, these accumulated defects become immobile or have very limited mobility. The difference in initial dislocation density with shock pressure in the $\alpha$ phase can be understood on the basis that reverse transformation occurs simultaneously during the shock loading, nucleating incipient $\alpha$ with a significantly lower density than the parent $\omega$. This supposition is supported for the initial EBSD analysis which shows much less lattice curvature in the newly transformed lenticular $\alpha$ grains post shock than in the equiaxed heavily deformed and twinned grains retained from the original microstructure. The difference in as-shocked dislocation density with shock pressure is likely due to a combination of the reverse transformation occurring to a lesser extent in the 10.5 GPa samples during the shock itself, a higher fraction of the heavily deformed $\alpha$ being retained from the initial microstructure and additional plastic deformation in the re-transformed $\alpha$ in the 10.5 GPa sample. Additional EBSD and other microscopy is required to deconvolve and individually evaluate these effects. 

Unlike the $\alpha$ phase, the initial dislocation density for the $\omega$ phase was approximately equal in both the 8 GPa and 10.5 GPa shocked samples (Figures \ref{fig:rms1}\subref*{fig:rms1_b} and \ref{fig:rms1}\subref*{fig:rms1_d}). It seems possible that the defect content in the $\omega$ saturates and does not further increase with increased peak pressure, however there is insufficient information to confirm this. Despite having similar initial dislocation densities in the $\omega$ phase for both shock pressures, the transformation rates and extent of transformation for different shock pressures were different. The most likely explanation is that the dislocation state of both samples are significantly different, thus requiring affirmation in a future study to determine the type and nature of dislocations present and their likely precursors in the pre-transformed $\alpha$. VISAR (Velocity Interferometer System for Any Reflector) analysis of the data from the shock experiments suggests that while both samples were above the transformation pressure for the same amount of time ($\approx 0.7 \mu s$) \cite{PhysRevB.89.220101}, the 8 GPa sample took almost this entire time to complete the forward transformation while the 10.5 GPa sample completed the forward transformation in $\approx 0.1\mu s$. Due to the more rapid nature of the shock transformation in the 10.5 GPa sample, the dislocations in the original $\alpha$ microstructure may have been ``caught" in the transformation and become sessile in the $\omega$ phase (stuck on planes with limited mobility). In contrast, slower transformation in the 8 GPa sample allowed more time for the $\alpha$ dislocations to move and react and avoid being trapped during shock transformation. Upon unloading, regions with dislocation densities lower than the critical dislocation density serve as nucleation sites for new $\alpha$, leaving behind $\omega$ regions with high dislocation density. It is likely that the slower moving transformation front in the 8 GPa samples pushed or swept dislocations in front of it, leaving larger regions of low defect density $\omega$ which could readily nucleate $\alpha$ immediately after the shock. An interesting thought to pursue in the future is that of the $\omega$ regions in the 8 GPa sample having more complex dislocation structures and locks than the 10.5 GPa sample due to the amount of time afforded for dislocation kinetics during the forward transformation during shock. Furthering this thought, although the average dislocation density of both differently shocked samples in the retained $\omega$ are closely similar and above a certain critical threshold, the 10.5 GPa sample would contain more regions with simpler structures and marginal stability which is more easily overcome during annealing. While speculative, this theory would account for the higher initial transformation rates at all temperatures and the higher $\alpha$ fraction saturation meta-stable state for the 10.5 GPa sample.  Confirmation of this hypothesis would require significant resources allocated to defect microscopy to determine the local dislocation state in the shock loaded material. 

The dependence of the hysteresis on thermal history and heating rate (see Figure \ref{fig:ramp}) further suggests that the mechanisms by which the arresting dislocations become mobile or are annihilated is complex or can take multiple paths. Samples heated at 1.5 K/s transformed to a more significant degree than the samples heated at 0.25 K/s, indicating that the various paths are not equivalent in a kinetic sense.  When heated at a slow rate, the defects may react by a variety of different pathways and possess a microstructure different to that of the as-shocked state when the sample finally reaches the target temperature. At higher temperature ramp rates, there is insufficient time for these initial low temperature reactions to occur and the dislocation structure at the target temperature becomes more representative of the as-shocked microstructure. As more energy becomes available at higher temperatures, it is likely that more complete dislocation reactions are possible. At slower ramp rate, significant dislocation debris or various locks may be formed with energy barriers which require even higher temperatures to overcome. 

%    One very surprising observation that supports this theory is that when the 8 GPa shocked sample was held at 503K for $\sim 3\times10^3$ seconds and subsequently raised to 773K, it did not entirely transform but retained $\sim 4\%$ $\omega$ phase (Figure \ref{fig:init_growth}). This is in contrast to a previous study by Brown et al., in which the material transformed completely by $\sim600$K when ramped at 0.05 K/s. It should be noted that during the 0.05 K/s ramp, the samples had been at temperatures in the transformation range (estimated as being above $443$K) for approximately $3\times 10^4$s or roughly an order of magnitude longer than the samples shown in Figure \ref{fig:alpha_evolution_3}. 

Figures \ref{fig:EBSD}\subref*{fig:EBSDf} and \ref{fig:EBSD}\subref*{fig:EBSDj} compare the $\alpha$ phase of 8 GPa peak pressure samples heated at 443K and 463K respectively. The samples at 443K did not show any visible evidence of nucleation of new $\alpha$, in contrast with the samples at 463K which displayed significant nucleation of lenticular $\alpha$ laths. Knowing that growth in $\alpha$ volume fraction occurs at both these temperatures, albeit with differing rates (refer to Figure \ref{fig:alpha_evolution_1}\subref*{fig:alpha_evolution_2_b}), it can be inferred that there is a higher energy barrier associated with nucleation in comparison to that of growth from existing $\alpha$. This is again counter to the predictions of Zong et al \cite{zong2014}, demonstrating the need for further in-situ experimental studies to inform the theory and modeling.  At sufficiently high temperatures, nucleation of $\alpha$ from $\omega$ parents dominates the reverse transformation, with the growth of the original $\alpha$ happening simultaneously. Unsurprisingly, the nucleation of $\alpha$ from $\omega$ parents occurs more readily near an existing $\alpha/\omega$ interface than in the bulk of a $\omega$ grain or $\omega/\omega$ grain boundary. However, in conjunction with the working hypothesis, this also means that mechanism responsible for the removal of arresting dislocations in the $\omega$ phase also occurs at similar sites. 

As a final mention, the results derived from EBSD analysis of the as-shocked microstructure reaffirmed the preferred orientation relationship initially observed by Song et al. \cite{song1995microscopic}. This orientation is maintained throughout growth of the $\alpha$ phase. In samples heated 463K and above, the newly nucleated lenticular $\alpha$ grains are quite similar to deformation twins in Zr, with respect to both their morphology and in the propensity to apparently initiate at interfaces and quickly propagate across the grain and terminate at either another $\omega$ grain boundary or another $\alpha$ lath. 

\section{Conclusions}
In-situ X-ray diffraction experiments were undertaken to better understand the stability and transformation behavior of $\alpha/\omega$ microstructure resulting from shock loading in pure Zr. Under standard conditions, the $\alpha$ phase is clearly the equilibrium phase. However, these experiments suggests that microstructure elements, presumably complex dislocation structures, arrest the reverse transformation and prevent the system from returning to equilibrium following shock loading, leading to samples with significant retained $\omega$ phase.

The current experiments quantitatively monitored that phase evolution during isothermal annealing and semiquantitatively followed the evolution of dislocation densities (via the evolution of RMS strain values) during isothermal annealing at 443, 463, 483, and 503K. At sufficiently high temperatures, the initial transformation occurred rapidly, quickly slowing down to reach a new metastable equilibrium state. Further, it was found that the samples shocked to higher peak pressures transformed at higher initial rates and asymptotically attained higher $\alpha$ fractions despite starting with significantly lower $\alpha$ in the as-shocked state.

The RMS strain was related to dislocation density and used to capture the evolution of microstructure of the $\alpha$ and $\omega$ phases in a semi-quantitative manner. The analysis suggests a high dislocation density in both the $\alpha$ and $\omega$ phases following shock loading. The dislocation density of the $\alpha$ phase drops significantly during heating, however we suggest that this is likely due to the growth of new $\alpha$ phase with low dislocation density rather than significant recovery or annealing in the $\alpha$ phase during heating. In contrast, there is an observed drop in dislocation density in the $\omega$ phase with a lesser extent compared to the $\alpha$ phase. It is postulated that the transformation is arrested due to complex dislocation structures in the $\omega$ phase ``pinning" or preventing the reverse transformation. This is supported by the observation that the reverse transformation initiates at a “critical” value of the RMS strain in the $\omega$ phase. As these complex dislocation structures are removed locally, the material proceeds to transform to $\alpha$. The relatively high concentration of defects in the $\omega$ phase following heating, along with the observation of a metastable state with retained $\omega$ reflects that the material is still prevented from transforming completely back to $\alpha$.

Some initial EBSD was performed to characterize the morphology of the as-shocked and partially annealed samples. The expected $\alpha/\omega$ orientation relationship is observed in the as-shocked state and this relationship is maintained throughout the transformation (growth and nucleation). Further, it was found that at 443K, only the growth of the existing $\alpha$ phase was observed while at 463K both nucleation of new $\alpha$ and growth of existing $\alpha$ grains were readily observed. This suggests that nucleation of new $\alpha$ grains requires overcoming a significant energy barrier or equivalently, the growth of existing $\alpha$ happens via a lower energy pathway than nucleation.

\section{Acknowledgements}
This work was supported by LDRD program funding at LANL. Los Alamos National Laboratory is operated by Los Alamos National Security LLC under DOE Contract DE-AC52-06NA25396. Use of the Advanced Photon Source, an Office of Science User Facility operated for the US DOE Office of Science by Argonne National Laboratory, was supported by the DOE under contract No. DE-AC02-06CH1135. SRN and BAW received additional support from the Center for the Accelerated Maturation of Materials, The Ohio State University, Columbus, OH 43210, USA. The authors would like to thank Ms. Laura Turcer and Ms. Elizabeth Krill for their assistance in preparing the samples for EBSD.

%Bibliography
\bibliographystyle{model1a-num-names}
\bibliography{biblio_list}

\end{document}